# Thermal Equilibrium Calorimeters
## — An Introduction

## Table of Contents

# Thermal Equilibrium Calorimeters — An Introduction

Dan McCammon

Physics Department, University of Wisconsin, Madison, WI 53706, USA

**Abstract.** Near-equilibrium thermal detectors operate as classical calorimeters, with energy deposition and internal equilibration times short compared to the thermal time constant of the device. Advances in fabrication techniques, cryogenics, and electronics have made it practical to measure deposited energy with unprecedented sensitivity and precision. In this chapter we discuss performance considerations for these devices, including optimal filtering and energy resolution calculations. We begin with the basic theory of simple equilibrium calorimeters with ideal resistive thermometers. This provides a starting point for a brief discussion of electrothermal feedback, other noise sources, various non-ideal effects, and nonlinearity. We then describe other types of thermometers and show how they fit into this theoretical framework and why they may require different optimizations and figures of merit. Most of this discussion is applicable also to power detectors, or bolometers, where the detector time constants may be short compared to variations in the incident signal power.

## 1 Introduction

Thermal detectors in general have a number of characteristics that make them more attractive than ionization detectors for many applications. Some of these are directly related to the lack of a requirement for efficient charge transport: if it is not necessary to collect electrons, large amounts impurities can be tolerated, and a radioactive source or specialized target material could be embedded within the detector. This also opens up a wide range of materials options, so the detector might in fact be made of the source. Other useful characteristics are sensitivity to exotic interactions that produce no ionization, and energy thresholds that can be made almost arbitrarily small.

Most detectors, including all ionization detectors, are non-equilibrium devices. In such a detector the idea is to get a substantial fraction of the deposited energy into a detection channel, and then to collect it as completely as possible before it decays into an undetectable channel. Here "channels" are the various forms that the internal energy of the detector can take, and the "detection channel" might for example be free charge (for an ionization detector), photons, phonons, or quasiparticles. To get good energy resolution, conditions must be made very uniform throughout the detection volume so that the branching ratio into the desired channel, channel lifetime, and collection efficiency are the same for all events. Even if this were done perfectly, however, statistical fluctuations in the branching still limit the energy resolution. In a typical ionization detector the charge channel gets only about 1/3 of the event energy, and statistical fluctuations in this fraction from event to event produce a fundamental constraint, or "Fano limit", on resolution [1]. For silicon, this is about 118 eV FWHM (full width at half maximum) at 6 keV.

Some thermal detectors also operate in a non-equilibrium mode, collecting only quasi-ballistic phonons or using sensors sensitive only to excitation energies >> k$T$. These can be fast relative to equilibrium devices, since thermal equilibrium often takes a very long time to establish at low temperatures, and with no restrictions on equilibration time they offer even more flexibility in choice of materials. Such detectors may suffer greatly from position dependence in the branching ratio into the detection channel and/or the lifetime and detection efficiency of the relevant excitations. They are also subject to branching statistics that are qualitatively the same as for ionization detectors. But for applications that require large volumes of dielectric material and do not need exceptionally good energy resolution, the speed advantage may outweigh other considerations.

Equilibrium detectors in principle offer the ultimate in energy resolution. With all channels in equilibrium, there are still fluctuations, but now many independent samples can be taken on a single event. For example, consider a thermometer embedded in a copper block. The energy content in the thermometer is the detection channel, while the electrons and phonons in the rest of the block contain most of the total energy. The energy content of the thermometer fluctuates, and the fluctuations can be fractionally large if the thermometer is very small and contains few excitations. However, if all the channels are well-coupled a very large number of independent samples are taken in a short time, and there is no fundamental limit on how accurately the total energy can be determined from the average energy density in the thermometer. In this chapter we will develop a quantitative statement of this argument and discuss the limits on resolution introduced by real thermometers and other departures from the ideal case. At this writing, the best resolution for thermal detectors is ~3 eV FWHM at 6 keV, so it is a worthwhile exercise to try to understand how far we can go with these devices.

## 2 Basic Linear Theory of Calorimeters

A simple calorimeter or bolometer has only three parts. As shown in Fig. 1, these are an absorber or thermal mass that contains the event or absorbs the incident power and thermalizes the energy, a perfectly coupled thermometer that measures the temperature increase of the absorber, and a weak thermal link to a heat sink that returns the absorber

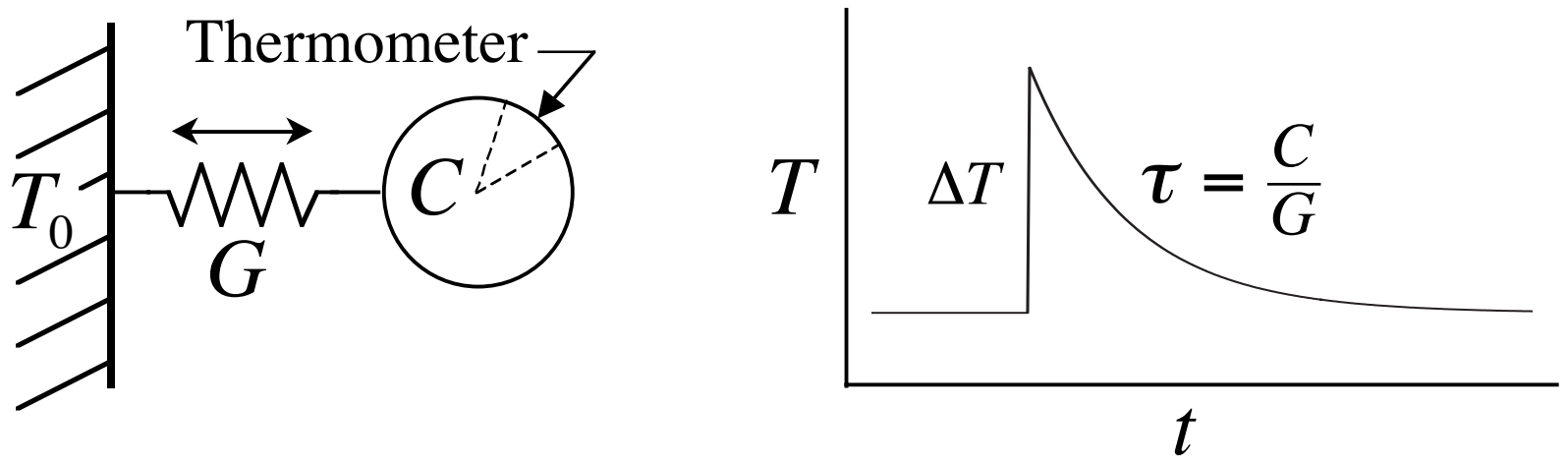

**Fig. 1.** The ideal calorimeter. An instantaneous energy input $E_0$ will raise the temperature by an amount $\Delta T = E_0/C$, and it will then decay back to its starting point with a time constant $\tau = C/G$

temperature to some defined value in the absence of a signal. The absorber can be characterized by its heat capacity *C*, the thermal link by its conductivity *G*, and the heat sink by its temperature $T_0$.

The same configuration can be used to measure a steady power input, *P*, with $\Delta T = P/G$. In this case the device is usually referred to as a bolometer. Such detectors have been used for many years to measure infrared radiation, and their theory is well-developed [2,3,4]. In particular, Mather has presented a complete linear theory for simple bolometers with ideal resistive thermometers [4] and made the straightforward extension to adapt these results to energy detectors [5]. These papers are somewhat terse. We will use a slightly different approach to arrive at the same results, and will keep much of the discussion of this section as general as possible so that it can be applied to detectors with all types of thermometers.

Section 2.1 contains a qualitative discussion of the most basic factors that influence energy resolution in any thermal equilibrium detector. It then uses thermometer Johnson noise as a basic example of resolution limited by a white noise source. Section 2.3 analyzes several major noise sources, of which only thermometer and load resistor Johnson noise are specific to resistive thermometers. The derivation in Sect. 2.4 of optimal filtering for energy detection should be entirely general for any linear system. Section 2.9 on the symmetry of equations for voltage and current output and 2.10 on common deviations from the simple detector of Fig. 1 are also quite general. The detailed derivation for resistive thermometers begins in Sect. 2.2. Optimization of the detector and bias power discussed in Sect. 2.5, which also contrasts the optimization of bolometers, or power detectors. Section 2.6 introduces circuit capacitance and inductance, 2.7 external feedback, and 2.8 the modifications necessary for thermometers where the resistance depends on voltage or current as well as temperature.

### 2.1 Limits on Energy Resolution: a first look

One irreducible source of noise comes from the random exchange of energy between the absorber and the heat sink over the thermal link. It is an elementary result of classical statistical mechanics that the magnitude of the resulting fluctuations in the energy content of the calorimeter is given by $\left\langle \Delta E^2 \right\rangle = k_B T^2 C$, independent of the conductance of the link [6]. If the energy carriers in the calorimeter have a mean energy $k_B T$, this can be thought of as Poisson fluctuations in their number.

These thermodynamic fluctuations represent a background against which the temperature increase due to an event must be measured, but they do not in themselves limit the accuracy of the measurement. The reason for this can be seen most easily in the time domain, as shown in Fig. 2b. This shows a simulation of a signal in the presence of thermodynamic fluctuation noise (TFN), and it is clear that the sudden increase in temperature due to the event can be measured quite precisely despite the large fluctuations if one looks closely enough at the "corners". This can be made quantitative in the frequency domain. Figure 3 shows the power spectrum of the exponential signal pulse and the noise power spectrum of the thermodynamic fluctuations. These have the same shape, with a single-pole roll off at $f_c = G/(2\pi C)$, so the signal-to-noise ratio is the same in all frequency bins. Each bin provides an estimate of the signal amplitude, and under rather general assumptions the noise in different bins is uncorrelated, so the signal to noise ratio will improve as the square root of the number of bins averaged. Averaging a bandwidth $\Delta f = G/C = 2\pi f_c$ will give a mean error just equal to the thermodynamic fluctuations, $\sqrt{k_B T^2 C}$, while $8\pi f_c$ will halve this value. By going to arbitrarily high frequencies, the signal can be measured to arbitrary accuracy.

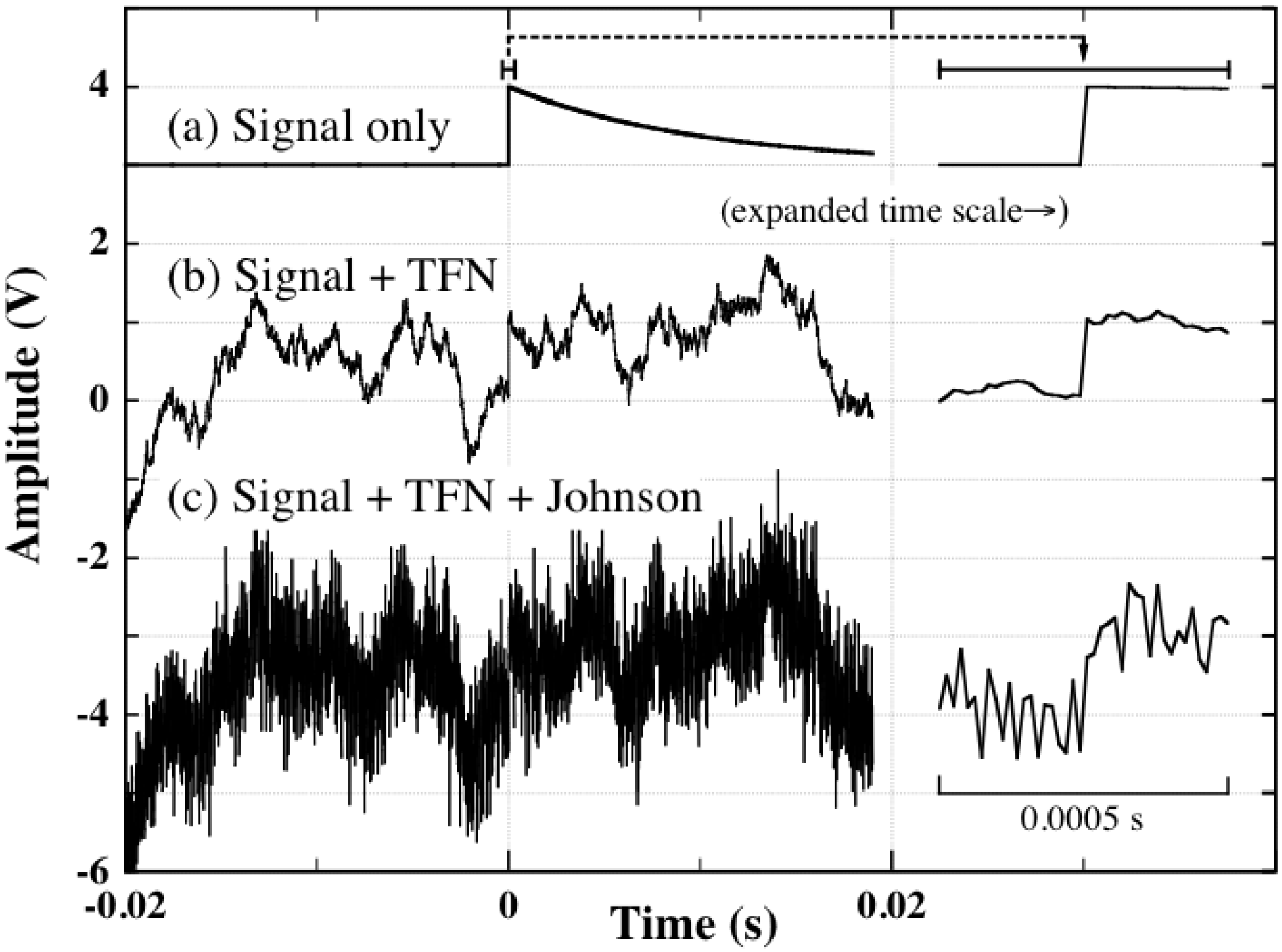

**Fig. 2.** Simulation of an event in an ideal calorimeter — time domain. Expanded views are shown at right. **(a)** Signal only, shows 10 ms thermal decay time constant. **(b)** Signal plus thermodynamic fluctuations as measured with a noiseless thermometer. The event energy can in principle be measured quite precisely, even though it is here just equal to the r.m.s. magnitude of the fluctuations. **(c)** Includes Johnson noise from a thermometer of sensitivity $\alpha \approx 250$ near optimum bias. The 50 kHz Nyquist frequency of the simulation was used as the noise bandwidth. This is ~3200 times $\tau_{\mathrm{sig}}^{-1}$.

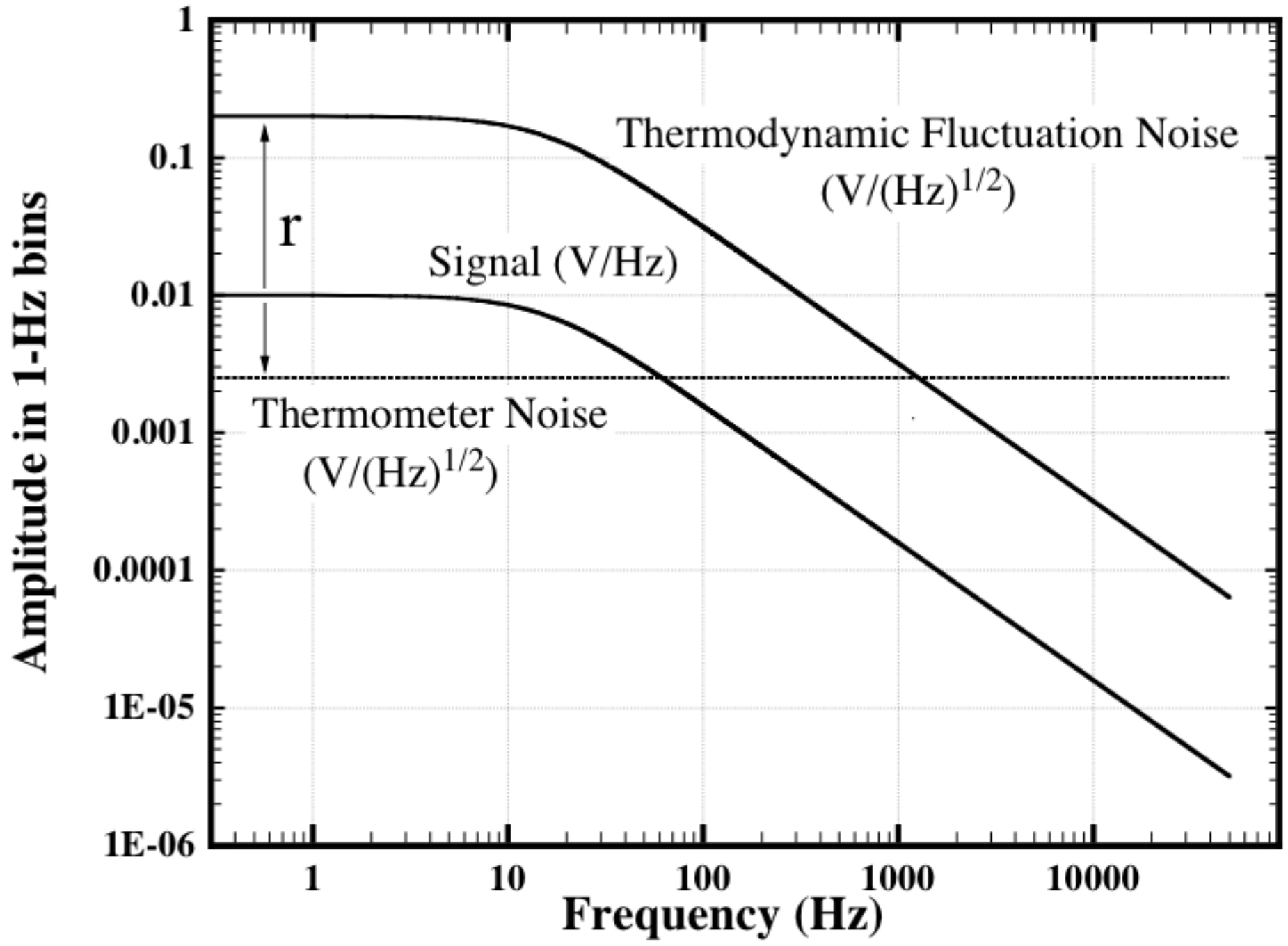

**Fig. 3.** Frequency domain. Note that the signal spectrum has different dimensions than the noise spectra.

There are a number of things that keep us from reaching this goal. Anything that makes the signal start falling faster than $1/f$ or makes the noise fall more slowly than this will produce a reduction in signal to noise ratio above the frequency — call it $f_{\text{bad}}$ — where this occurs. The resolution improves as the square root of the bandwidth as long as the s/n ratio is constant, but little more is gained once it starts dropping rapidly, so the "useful bandwidth" will be proportional to $f_{\text{bad}}$. If energy is deposited or thermalized over a finite time, or if the absorber is imperfectly coupled to the thermometer, the signal will have a finite risetime and its power spectrum will have an additional pole above which it falls off as $1/f^2$. This limitation is discussed in more detail in Sect. 2.10. Alternatively, there might be some additional noise source, such as amplifier noise, that is independent of frequency. Since the thermodynamic fluctuation noise is dropping as $1/f$, it will at some point reach this "noise floor," and the total noise spectrum will flatten out. The energy uncertainty is proportional to $(f_c/f_{\text{bad}})^{1/2}$ and to the magnitude of the TFN, $\sqrt{k_B T^2 C}$, so for a high-resolution detector one would like to minimize both.

In the sections below, we analyze in detail the operation of detectors with an ideal resistive thermometer, or thermistor. This important class includes both the standard doped semiconductor thermometer and the promising superconducting transition edge sensor (TES). A thermistor might seem to be a poor choice for a calorimeter, first because it has an irreducible Johnson noise and second because it transduces temperature changes as resistance changes which then require power dissipation to read out. This readout power will warm the detector and increase the energy fluctuations. We will see that the very high sensitivity of available thermistors does much to mitigate these drawbacks, and they are currently the most common thermometer type. Here we use the Johnson noise of such a thermometer as a concrete example of how a white noise source limits the useful bandwidth, and of the optimal operation of a detector where the thermometer signal to noise ratio increases with readout power.

The effect of the Johnson noise is shown in Fig. 2c. At frequencies above the point where it crosses the thermodynamic fluctuation noise, the signal-to-noise ratio per frequency bin starts dropping linearly with $f$ as can be seen in Fig. 3, and there is less and less gain from including higher-frequency bins in the total. The useful bandwidth is proportional to $r$, the ratio of the low-frequency TFN to the Johnson noise, since the TFN $\propto 1/f$ above $f_c$, making the crossing point $r \cdot f_c$. The Johnson noise is fixed in voltage spectral density, while the TFN is a temperature spectral density, so $r$ depends on both the thermometer sensitivity in converting temperature to resistance changes and on the readout current that converts these to voltage changes. It is convenient to define a dimensionless local sensitivity for the thermometer

$$\alpha \equiv \frac{d\log R}{d\log T} = \frac{T}{R}\frac{dR}{dT}, \tag{1}$$

and the readout current can be parameterized by the temperature rise $\Delta T_{\rm Bias}$ it causes when the dissipated power flows to the sink over the thermal link. For a very small temperature rise, one then gets the simple result that $r \approx \alpha(\Delta T_{\rm Bias}/T_0)^{1/2}$, where $T_0$ is the heat sink temperature.

The temperature increase clearly has an optimum value. If the bias current and $\Delta T_{\rm Bias}$ are very small, then the signal and transduced TFN are also very small in terms of voltage and will be completely dominated by the Johnson noise. On the other hand, a large $\Delta T_{\rm Bias}$ significantly increases both *T* and *C*, greatly increasing the TFN. The optimum value of $\Delta T_{\rm Bias}/T$ depends on the temperature coefficients of *C* and *G*, and on $\alpha$ for $\alpha < \sim 5$. For practical devices this value is between 0.11 and 0.20 [5].

We can anticipate some results below and express the minimum energy uncertainty due to these two noise sources alone as $\Delta E = \xi\sqrt{k_B T_0^2 C_0}$, where $C_0$ is the heat capacity at the heat sink temperature (see (37)). With $\Delta T_{\rm Bias}$ optimized, the only strong dependence of $\xi$ is on $\alpha$. Since the signal-to-noise ratio scales with the square root of the usable bandwidth, which is in turn proportional to *r* and $\alpha$, we expect $\xi$ to scale as $\alpha^{-1/2}$ for large values of $\alpha$. In fact, $\xi \approx 5\alpha^{-1/2}$, so the energy resolution can be much better than the magnitude of the thermodynamic fluctuations for very high thermometer sensitivities. Inspection of Fig. 3 shows that achieving this result depends on meeting the stringent requirement that there not be another pole in the signal response (due to thermalization time or internal time constants in the detector) at a frequency below about $\alpha f_c$. This is discussed further in Sect. 2.10. Figure 3 also shows that for a very sensitive thermometer, the Johnson noise will appear negligible compared to the TFN at low frequencies, yet the energy resolution is still entirely dependent on its level..

We will now look more quantitatively at the detector response and noise, both to justify the assertions above and to include additional effects. These are first derived in terms of voltage output, but see Sect. 2.9 for conversion to current output.

## 2.2 Power responsivity as function of frequency

This derivation is done in steps in an attempt to keep it as transparent as possible. I have also introduced several definitions that are more or less commonly used to keep later equations from becoming too dense. These are by no means standardized, and they have been borrowed without attribution from a number of sources based variously on a) breadth of use, b) logical consistency, or c) earliest definition. For those new to thermal calculations, note that most of the familiar solutions to electrical circuits can be taken over using the analogies for thermal variables shown in Table 1.

For historical reasons, the derivations of this section are done with the assumption that the output signal is the voltage across the thermistor. As explained in part 2.9, however, the symmetry of Kierkhoff's circuit equations results in a very simple transformation that will convert these to the current-output form. As a result, for example, the current output of a voltage-biased positive temperature coefficient detector is identical to the voltage output of a current-biased negative temperature coefficient detector. Table 5 gives the current-out form for all of the equations derived in this section. If you have been working primarily with current output (as is usual for TES detectors), you are encouraged to skip ahead and read Sect. 2.9 before continuing.

**Table 1.** Thermal–electrical analogies

| Basic Quantities | | | Equivalent Equations |
|---|---|---|---|
| temperature: | $T \rightarrow V$ | (voltage) | $T = P/G \rightarrow V = IR$ |
| thermal power: | $P \rightarrow I$ | (current) | $dT/dt = P/C \rightarrow dV/dt = I/C$ |
| thermal conductivity: | $G \rightarrow 1/R$ | (conductance) | |
| thermal energy: | $E \rightarrow Q$ | (charge) | |
| heat capacity: | $C \rightarrow C$ | (capacitance) | |

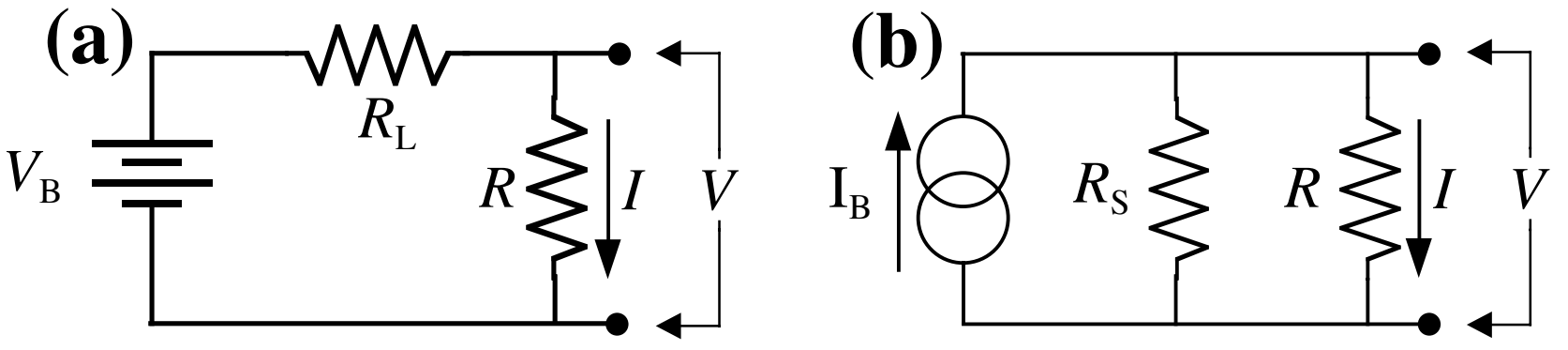

**Fig. 4.** Common detector bias circuits: (**a**) bias with series load impedance, $R_L$, and (**b**) a circuit often used when $R_L$ is small. Since (b) is readily converted to (a) by Thevenin's theorem ($V_B = I_B R_S$, $R_L = R_S$), we will consider only (a)

***D.C. response.*** We can write the change in voltage across a resistive thermometer for a given power input as

$$\frac{dV}{dP_{in}} = \frac{dT}{dP_{in}} \frac{dR}{dT} \frac{dV}{dR}. \tag{2}$$

A steady energy input $P_{in}$ will produce a temperature rise in the absorber $\Delta T = P_{in}/G$ as it flows through the thermal link to the head sink, so $dT/dP_{in} = 1/G$. Using the definition (1) for thermometer sensitivity $\alpha$, $dR/dT = \alpha R/T$. For the bias circuit shown in Fig. 4a, $dV/dI = -R_L$ and we can write for the voltage across the thermistor

$$V = IR \quad \Rightarrow \quad \frac{dV}{dR} = I + R\frac{dI}{dV}\frac{dV}{dR} \quad \Rightarrow \quad \frac{dV}{dR} = \frac{R_L}{R + R_L} I = K_L I, \tag{3}$$

where for future convenience we define $K_L \equiv R_L/(R+R_L)$. For $R_L >> R$, $K_L = 1$. For lower values of $R_L$, $K_L$ represents the "loading factor" of the load resistance. Substituting these quantities into (2) and writing the bias power dissipated in the thermometer as $P = I^2 R$, we get

$$\frac{dV}{dP_{in}} = \frac{1}{G}\frac{R}{T}\alpha K_L I = \frac{\alpha P}{GT}\frac{1}{I}K_L = \frac{L_0}{I}K_L. \tag{4}$$

The quantity $L_0 \equiv \alpha P/(GT)$ appears often in calorimeter calculations, and can be regarded as the dimensionless D.C. "gain".

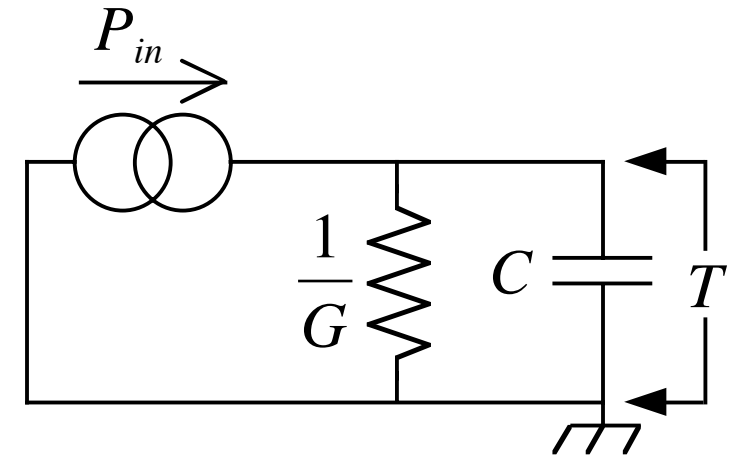

**Fig. 5.** Thermal equivalent circuit

***A.C. response.*** Figure 5 shows the thermal circuit equivalent of the calorimeter of Fig. 1. The net current onto the capacitor is $P_{in} - GT$, so we can write

$$\frac{dT}{dt} = \frac{1}{C}(P_{in} - GT). \tag{5}$$

For $P_{in}$ a delta-function, the solution to (5) is clearly just the exponentially decaying pulse shown in Fig. 1. If we assume $P_{in}(t) = P_0 e^{i\omega t}$, then the solution is

$$T(t) = \frac{P_0 e^{i\omega t}}{G(1+i\omega\tau)}, \tag{6}$$

where $\tau \equiv C/G$. We can then write

$$\frac{dT}{dP_{in}} = \frac{1}{G}\frac{1}{(1+i\omega\tau)}, \tag{7}$$

and using this in (2) makes (4) become

$$\frac{dV}{dP_{in}} = \frac{L_0}{I(1+i\omega\tau)}K_L \equiv A(\omega) \quad \text{(volts/watt)}. \tag{8}$$

***Electrothermal feedback.*** Up to this point we have been ignoring the fact that the bias power $P = I^2 R = V^2/R$ is part of $P_{in}$. We can regard the bolometer as an "amplifier" with a thermal power input and an electrical voltage output. The "gain" of this circuit is then given by (8). The contribution to $P_{in}$ due to V can be regarded as "feedback". Going back to the purely electrical case as shown in Fig. 6, if $\beta \equiv \partial V_{\text{in}}/\partial V_{\text{out}})_{V_{\text{ext}}}$, then $V_{in} = V_{ext} + \beta V$, and $V = AV_{in} = A(V_{ext} + \beta V)$. This leads to the standard result for the "closed-loop gain":

$$A_{CL} \equiv \frac{V}{V_{ext}} = A\frac{1}{1-\beta A}, \tag{9}$$

where the product $\beta A$ is called the "loop gain".

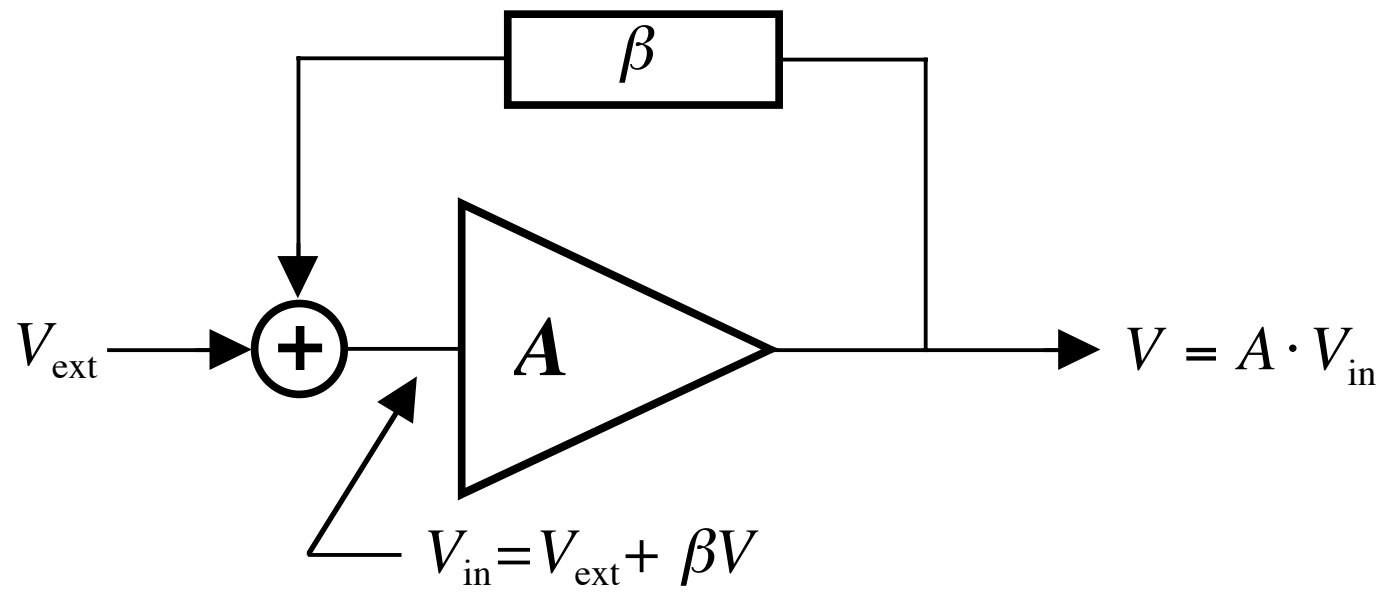

**Fig. 6.** Feedback circuit

For our bolometer, the input to $A$ is thermal power, and $V_{\text{ext}}$ becomes $P_{\text{ext}}$. The output is still voltage, so $A$ and $\beta$ now have dimensions, although their product necessarily does not. Now $A$ is given by (8) and $\beta = dP/dV$ is the change in bias power into the detector caused by a change in voltage at the output, or "electrothermal feedback" (ETF). With $P = IV$, we have $dP = IdV + VdI$. From the bias circuit in Fig. 4a, $dI/dV = -1/R_L$ and

$$\beta = \frac{dP}{dV} = I + V\left(\frac{-1}{R_L}\right) = I\left(1 - \frac{R}{R_L}\right). \tag{10}$$

Taking the gain $A$ from (8), the loop gain is then

$$\beta A = I\left(\frac{R_L - R}{R_L}\right)\cdot\left(\frac{L_0}{I(1+i\omega\tau)}\right)K_L = \left(\frac{R_L - R}{R_L + R}\right)\frac{L_0}{(1+i\omega\tau)} = \frac{bL_0}{(1+i\omega\tau)}, \tag{11}$$

where we have defined

$$b \equiv \frac{R_L - R}{R_L + R} = 2K_L - 1. \tag{12}$$

Our responsivity including electrothermal feedback (the "closed loop gain") is now

$$S_V(\omega) \equiv \frac{dV}{dP_{ext}} = A_{CL} = A\frac{1}{1-\beta A} = AK_F, \tag{13}$$

where $K_F$ is the "feedback factor" giving the change in gain introduced by ETF. Therefore

$$K_F \equiv \frac{1}{1-\beta A} = \frac{1}{1 - bL_0/(1+i\omega\tau)} = \frac{1}{1-bL_0}\cdot\frac{1+i\omega\tau}{1+i\omega\tau/(1-bL_0)}, \tag{14}$$

and

$$S_V(\omega) = \frac{L_0}{I}\frac{1}{1+i\omega\tau}K_L K_F = \frac{L_0}{I(1-bL_0)}\frac{1}{1+i\omega\tau_e}K_L \quad \text{(V/W)}, \tag{15}$$

where $\tau_e \equiv \tau/(1-bL_0)$. Note that (12) shows the ETF is controlled entirely by the load resistance, since $b$ and the resulting feedback depend on the value of $R_L$ relative to $R$. For a negative temperature coefficient thermometer ($\alpha$ and $L_0 < 0$), one gets negative feedback for $R_L > R$ and positive feedback for $R_L < R$. For a positive temperature coefficient, the reverse is true. In both cases, there is *no* electrothermal feedback when $R_L = R$.

We will point out in Sect. 2.5 that in the idealized linear case we are considering here electrothermal feedback has no effect on the *NEP*, energy resolution, or count rate capability. However, for real detector systems there can be large practical benefits. Positive ETF has been used to boost signal levels to reduce the effect of amplifier noise, but the biggest benefits come from exploiting the large negative ETF available with very high sensitivity thermistors. This can stabilize gain and reduce nonlinearity effects, as first discussed extensively by Irwin [7]. One can get some idea of the importance of this for high-$\alpha$ detectors by trying to imagine using a high gain operational amplifier without negative feedback. Flattening the gain over a wide bandwidth and stabilizing the operating point are two major benefits, although there are differences of degree from the op-amp case. The nonlinearities are worse, the loop gain $\beta A$ usually does not have the very large values needed to stabilize the gain at $1/\beta$ to the desired accuracy, and $\beta$ itself depends on a varying $I$ and on $R$ for finite values of $R_L$. Section 3 discusses consequences of large signals, where single events heat the detector enough to change its properties and alter the response to additional events that occur before the detector has returned to its equilibrium temperature. Negative ETF shortens the cooling time by a factor of $1/(1-bL_0)$, and can greatly reduce response variability between events at high count rates when such large signals are present.

## 2.3 Major noise terms

Now that we have the voltage responsivity of the detector to an arbitrary signal power input, we need to find the noise voltage at the output in order to determine how precisely signals can be measured. We calculate the output noise voltage below for the most important noise terms. These are generally uncorrelated, so the total output noise can be found by simply adding their squares. A common figure of merit for power detectors is the r.m.s. power required at the input at a given frequency to produce an output voltage equal to the r.m.s. noise voltage in a unit bandwidth at that frequency. This noise equivalent power, or *NEP*, is given by $e_n(\omega)/S_V(\omega)$.

***Thermodynamic fluctuation noise.*** One unavoidable source of noise is the statistical fluctuations in the energy content of the detector produced as it exchanges energy with the heat sink. One can derive directly from fundamental assumptions and definitions of statistical mechanics that

$$\langle \Delta E^2 \rangle = k_B T^2 C, \tag{16}$$

where $C$ is the detector heat capacity [6]. This, however, says nothing about their frequency spectrum. The fluctuations are produced by a noise power flow over the thermal link $G$, and the power in a unit bandwidth at frequency $\omega$ could be represented by the $P_{in}$ of Fig. 5, resulting in temperature fluctuations given by (6). If we assume for now that this power spectrum is a shot noise and therefore independent of frequency, then $P_0$ is constant. As shown, for example, by Richards [7] this can be integrated over all frequencies to give the total energy fluctuation (16), which requires the spectral density of $P_{in}$ to be

$$p_{\text{TFN}}^2 = 4k_B T^2 G \ \ \text{W}^2/\text{Hz}. \tag{17}$$

A white spectrum is the only fixed spectrum that will give the correct result (16) for an arbitrary choice of $C$, and is also the result of detailed calculations of the power flow on the link for cases with simple physics.

Equations (16) and (17) are valid only in thermal equilibrium, where the temperature $T$ of the detector is equal to the heat sink temperature $T_0$. In general, the detector will be at some higher temperature, most usually due to the bias power $P$ used to read out the thermometer. The power spectral density in the link will then depend on details of the nature of the link. Two limiting cases have been worked out. When the mean free path of the energy carriers is large compared to the length of the link (the radiative or specular limit), Boyle & Rogers [9] find

$$p_{\text{TFN}}^2 = 4k_B T_0^2 G_0 \cdot \frac{t^{\beta+2}+1}{2} \ \ (\text{W}^2/\text{Hz}), \tag{18}$$

while for mean free path small compared to the length (diffusive limit) Mather [4] gets

$$p_{\text{TFN}}^2 = 4k_B T_0^2 G_0 \cdot \frac{\beta+1}{2\beta+3} \cdot \frac{t^{2\beta+3}-1}{t^{\beta+1}-1} \ \ (\text{W}^2/\text{Hz}), \tag{19}$$

where $t \equiv T/T_0$ (here and for the remainder of the chapter), and $T^{\beta}$ is the assumed temperature dependence of $G$. Note that we are using the most conventional definition for the link thermal conductivity, $G_{\text{END}} \equiv \partial P_{\text{INTO LINK}}/\partial T_{\text{END}}$, which is different at the two ends of a link with a temperature gradient, but is always positive. How this $G$ is related to the thermal conductivity constant is shown nicely in [4]. In this chapter, $G$ without a subscript refers to the detector or hot end. Equations (18) and (19) have been normalized to $T$ and $G$ at the heat sink temperature for later convenience in optimizing the value of $t$. We also collect the thus-normalized link properties into functions $\text{F}_{\text{LINK}}(t,\beta,\text{link physics})$ such that (18) and (19) can both be written as

$$p_{\text{TFN}}^2 = 4k_B T_0^2 G_0 \cdot \text{F}_{\text{LINK}}(t,\beta) \tag{20}$$

Since $p_{\text{TFN}}$ is equivalent to an external power input $P_{\text{ext}}(\omega)$ in each frequency interval, we can use (15) to write the output noise spectral density as

$$e_n(\omega)_{\text{TFN}} = p_{\text{TFN}} \cdot S_V(\omega) = \left(4k_B T_0^2 G_0 \cdot \text{F}_{\text{LINK}}(t,\beta)\right)^{1/2} \cdot \frac{L_0}{I} \frac{1}{1+i\omega\tau} K_L K_F \ \ \left(\text{V}/\sqrt{\text{Hz}}\right). \tag{21}$$

***Thermometer Johnson noise.*** Another irreducible noise source for detectors with resistive thermometers is the Johnson or Nyquist noise in the thermometer. This can be modeled as a voltage source with spectral density $e_{nJ}\ V/\sqrt{Hz}$ in series with a noiseless (but temperature sensitive) resistance R, where $e_{nJ} = (4k_B TR)^{1/2}$. In the bias circuit of Fig. 4a, the output voltage in a unit bandwidth would be $dV = e_{nJ} \cdot R_L/(R_L + R) = e_{nJ} K_L$ if there were no thermal effects. However, the bias current $I$ does work $Ie_{nJ}$ on the Johnson noise source. This is physically located in the detector and therefore heats it and changes the resistance. The total power is $IV$, so the expression for $\beta \equiv \partial P_{in}/\partial V$ in (10) is still correct to first order in $e_{nJ}$, and we can use the equivalent circuit shown in Fig. 7. We can then write $dV = e_{nJ} K_L + A\beta \cdot dV$, which can be solved for the output noise spectral density due to the Johnson noise:

$$e_{nJ\text{-Therm}}(\omega) = dV = e_{nJ} \cdot \frac{K_L}{1-\beta A} = \sqrt{4k_B TR} \cdot K_L K_F. \qquad (22)$$

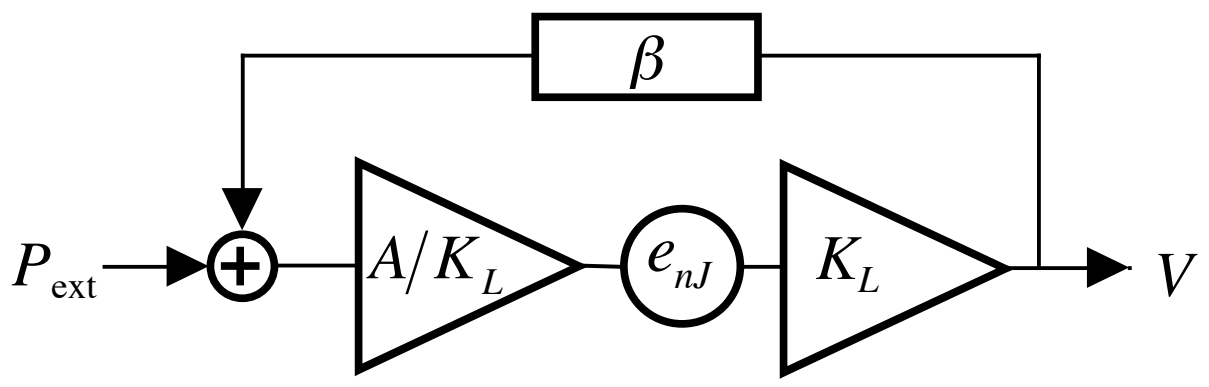

**Fig. 7.** Johnson noise source equivalent circuit

The same result can be obtained by direct differentiation of $V = IR + e_{nJ}$, noting that $dR/de_{nJ} = (dR/dT)(dT/dP)(dP/de_{nJ})$, with $P = IV$ and $dI = -dV/R_L$, and retaining only first-order terms in $e_{nJ}$.

***Load resistor Johnson noise.*** Referring to Fig. 4a, we can model the load resistor Johnson noise as a voltage source with spectral density $e_{nJRL} = \sqrt{4k_B T_L R_L}$, where $T_L$ is the physical temperature of the load resistor. Following the same approach used above for detector Johnson noise, we first assume no thermal effects. We then have a simple voltage divider, and the change in detector voltage $V$ produced by $e_{nJRL}$ is $dV = e_{nJRL} \cdot R/(R + R_L)$. We write this as $e_{nJRL} \cdot (R/R_L)K_L$, so we can simply substitute $e_{nJRL} \cdot R/R_L$ for $e_{nJ}$ in the circuit of Fig. 7.

Now, however, the expression for $I$ in (10) must be modified to $I = (V_B + e_{nJRL} - V)/R_L$. This gives rise to an extra term in $dP$, which becomes $dP = \beta dV + V\, e_{nJRL}/R_L$. The $\beta dV$ term is still taken care of by the feedback loop, and we can represent the new term as "$P_{\text{ext}}$" $= e_{nJRL} \cdot IR/R_L$ in Fig. 7. The circuit equation is then $V = e_{nJRL}(R/R_L)K_L + A(\beta V + e_{nJRL}\, IR/R_L)$. Solving for $V$ and using $A$ from (8) and $K_F = 1/(1-\beta A)$:

$$e_{nJ\text{-Load}}(\omega) = V = \frac{K_L + AI}{1-\beta A}\left(\frac{R}{R_L}\right) e_{nJRL} = \frac{R}{R_L}\left(1 + \frac{L_0}{1 + i\omega\tau}\right) K_L K_F \cdot \sqrt{4k_B T_L R_L}. \qquad (23)$$

At the cost of considerably more algebra, this can also be obtained by direct evaluation of the derivatives in $dV/de_{nJRL} = (dV/dI)(dI/de_{nJRL})$. It is worthwhile to calculate the first of these, since $dV/dI$ is the dynamic impedance $Z(\omega)$ of the detector, a function that is easily measured and provides a valuable experimental diagnostic of detector characteristics [10]. The usual approach is to differentiate $V = IR$, writing $dR/dI = (dR/dT)(dT/dP)(dP/dI)$ with $dT/dP$ from (7) and differentiating $P = I^2R$ to get $dP/dI$. Substituting and solving for $dV/dI$ gives

$$Z(\omega) \equiv \frac{dV}{dI} = \frac{1 + L_0 + i\omega\tau}{1 - L_0 + i\omega\tau} \cdot R \tag{24}$$

The response to load resistor noise is then just the voltage-divider formed by $Z_L$ and $Z(\omega)$: $e_n(\omega) = e_{nJRL} \cdot Z(\omega)/(Z(\omega) + R_L)$. With some manipulation, this gives the result (23). The load resistor Johnson noise can normally be made negligible by choosing $R_L/R >> 1$, but care must be taken to chose a resistor type that does not produce large amounts of "excess noise" when the bias current flows through it.

***Amplifier noise.*** Amplifier noise is normally specified as a voltage noise source $e_{nA}$ in series with the amplifier input and a current noise source $i_{nA}$ in parallel with the input as shown in Fig. 8. The voltage noise then simply adds to the noise at the detector. If we had modeled the load resistor Johnson noise as its Thevenin-equivalent current source in parallel with $R_L$, we would see that the amplifier current noise source occupies effectively the same position (the battery can be shorted as far as changing signals are concerned). So we use the responsivity from (23) for this, and the total noise from the amplifier referred to the voltages at its input is

$$e_{n\text{-AMP}}(\omega) = \sqrt{e_{nA}^2(\omega) + i_{nA}^2(\omega) \cdot \left( R\left(1 + \frac{L_0}{1 + i\omega\tau}\right) K_L K_F \right)^2} \, . \tag{25}$$

This assumes that the amplifier current and voltage noises are uncorrelated, which may not be the case.

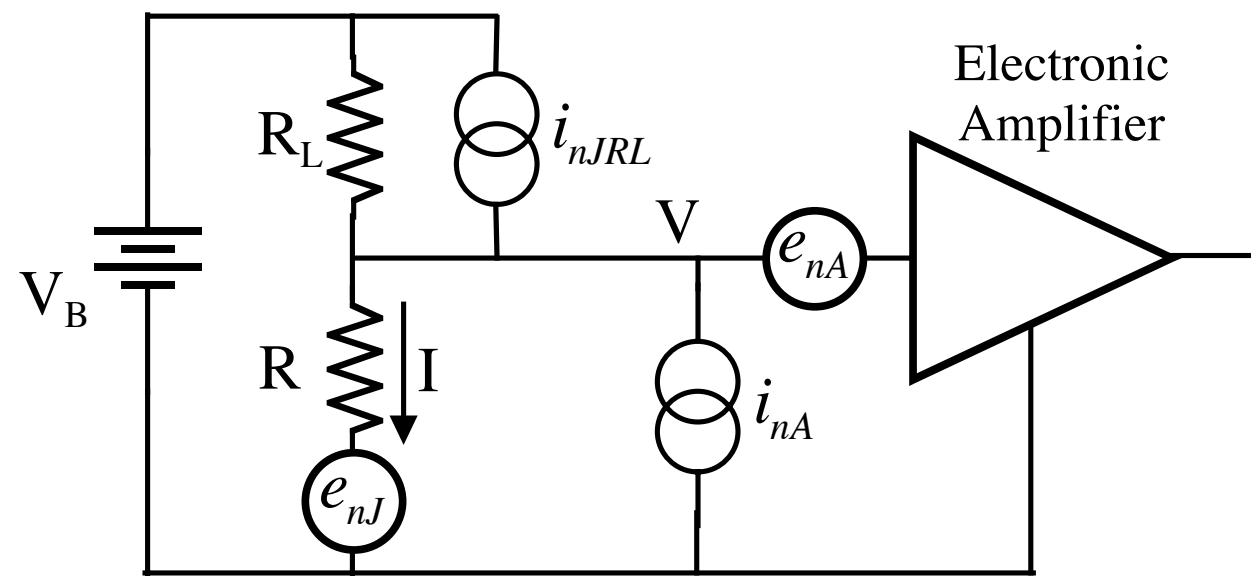

**Fig. 8.** Equivalent circuit showing definitions of amplifier noise and Johnson noise sources

It is usually possible to make the amplifier noise negligible for the commonly-used thermistor types. Doped semiconductor thermometers are easily made with resistances of tens of megohms. At this source resistance, silicon junction field effect transistors (JFETs) operated at ~120 K have noise temperatures less than 5 mK, far below the operating point of today's detectors. Superconducting transition edge sensors (TESs) typically have resistances of less than one ohm, where superconducting quantum interference device (SQUID) amplifiers are even better than JFETs are at high impedance. Multiplexing multiple detectors onto an amplifier eats into this margin however, and amplifier noise can become a major consideration for such schemes.

***Photon background noise.*** Photon background absorbed by a detector has two effects. First, it is an power source that raises the temperature of the detector while adding no readout power. This degrades the detector performance and requires a different optimization, as discussed for example in [11]. Second, the quantized photon energy produces shot noise, which appears as a noise power at the input to the detector in the same way that TFN does. The input noise power can be calculated by integrating the shot noise over the photon spectrum. For high-efficiency systems and $h\nu << k_B T$ it is necessary to take into account the Bose statistics that greatly increase the fluctuation level. A useful form of this calculation is given in [4], while a more extensive treatment can be found in [12]. This background is often very important for infrared detectors, but for calorimeters detecting higher energy photons it is generally assumed that it can be made negligible through proper optical filtering. In practice, it is much more difficult that it seems to actually achieve this, and "light leaks" should be one of the suspects when unexpected noise is observed.

***Additional thermometer noise.*** All resistive thermometers must have Johnson noise, but they may have additional noise sources as well. These are specific to the thermometer type, and are discussed in the following chapters. Their contribution to the output voltage can usually be treated in the same way that the Johnson noise was above. For instance, thermometer resistance fluctuations $r_n$ can be multiplied by *I* to give a voltage fluctuation that is substituted for $e_{nJ}$ in (22). This gives

$$e_{n-\Delta R} = K_F K_L I \cdot r_n, \text{ or } \left\langle \frac{\Delta V^2}{V^2} \right\rangle = \left(K_F K_L\right)^2 \left\langle \frac{\Delta R^2}{R^2} \right\rangle. \tag{26}$$

## 2.4 Optimal Filtering and Energy Resolution

We assume the signal is of the form $V(t) = E_0 f(t)$, where $f(t)$ is known and independent of $E_0$. (Note: In this subsection, *t* is time.) The problem is to extract the best estimate of $E_0$ in the presence of a known random noise of spectral density $e_n(f)$. This problem can be solved rather generally for a linear system, which we have assumed in the analysis of this section. The calculation is difficult in the time domain because the noise in different time bins is correlated if its spectrum is frequency dependent. In the frequency domain, however, the noise in different bins is uncorrelated on the condition that it is stationary, which means that its statistical properties do not change during the signal pulse. This seems likely to hold for almost any system in the linear small-signal limit. (Filtering of large signals is discussed in Sect. 3.)

We take the discrete Fourier transform of V(t) to obtain the values of $s_i$, which is the signal amplitude in the $i^{th}$ frequency bin. The root mean square value of the noise voltage in a bin is given by $n_i$. Each $s_i$ is proportional to $E_0$, so if the noise is uncorrelated, every bin provides an independent estimate of its magnitude. We can choose a set of weights $w_i$ and combine all the bins to get an expected value for the signal of $E = \sum_{i=1}^{\infty} w_i s_i$, and a corresponding noise fluctuation $\Delta E_{\text{rms}} = \left( \sum_{i=1}^{\infty} (w_i n_i)^2 \right)^{1/2}$. We want choose $w_i$ to maximize $E/\Delta E_{\text{rms}}$, so we take the derivative of this ratio with respect to an arbitrary $w_k$ and set it equal to zero, giving

$$w_k = \frac{s_k}{n_k^2} \left( \sum_{i=1}^{\infty} (w_i n_i)^2 \Big/ \sum_{i=1}^{\infty} w_i s_i \right). \tag{27}$$

The ratio $E/\Delta E_{\text{rms}}$ is clearly independent of any common scale factor on the $w_i$, so we simply drop the constant term in brackets. The $s_i$ are in general complex. The denominator of $E/\Delta E_{\text{rms}}$ depends only on the absolute value of the $w_i$, so we are free to choose their phases to maximize the numerator. This is accomplished by making each term entirely real, with

$$w_i = \frac{\hat{s}_i}{n_i^2}, \tag{28}$$

where $\hat{s}_i$ is the complex conjugate of $s_i$. This will make $E$ a pure cosine sum, and in the time domain the filtered signal $E(t)$ will always peak at $t = 0$.

To get the resulting energy resolution, we must be careful with the normalization of the Fourier transforms. We will use the pair

$$H(t) = \int_{-\infty}^{\infty} h(f) e^{-i2\pi f t} df \quad \text{and} \quad h(f) = \int_{-\infty}^{\infty} H(t) e^{i2\pi f t} dt \ . \tag{29}$$

We assume a power input to the detector $E_0 P(t)$, with Fourier transform

$$\int_{-\infty}^{\infty} E_0 P(t) e^{i2\pi f t} dt = E_0 p(f). \tag{30}$$

Using the detector responsivity (15), gives the detector output voltage spectrum $v(f) = E_0 S_V(f) p(f) = E_0 s(f)$. The optimal filter in (28) is $\hat{s}(f)/e_n^2(f)$, so the filtered signal becomes $v_{\text{FILT}}(f) = E_0\, \hat{s}(f) s(f)/e_n^2(f)$. Transforming this to the time domain,

$$V_{\text{FILT}}(t) = E_0 \int_{-\infty}^{\infty} \frac{\hat{s}(f) s(f)}{e_n^2(f)} e^{-i2\pi f t} df, \quad \text{and} \quad V_{\text{FILT}}(0) = 2E_0 \int_0^{\infty} \frac{|s(f)|^2}{e_n^2(f)} df \ , \tag{31}$$

where in the second integral we have made use of the fact that $s(f)$ is the transform of a real function of *t*, so $|s(-f)|^2 = |s(f)|^2$. $V_{\text{FILT}}(0)$ is our best estimate of the input signal amplitude. (In practice, the filter coefficients might have an arbitrary normalization, and the energy scale would be determined empirically from events of known energy.)

The filtered noise is $e_{n\text{-FILT}}(f) = \hat{s}(f)e_n(f)/e_n^2(f) = \hat{s}(f)/e_n(f)$. This is uncorrelated at different frequencies, so we can sum its absolute square and get

$$\left\langle V_{n\text{-FILT}}^2 \right\rangle = \int_0^\infty \frac{|s(f)|^2}{e_n^2(f)} df, \tag{32}$$

the mean square fluctuation expected at any time, including $t = 0$. We can normalize it to energy units by dividing by $V_{\text{FILT}}^2(0)$ from (31) with $E_0$ equal to one energy unit:

$$\left\langle \Delta E^2 \right\rangle = \frac{\int_0^\infty \frac{|s(f)|^2}{e_n^2(f)} df}{\left(2\int_0^\infty \frac{|s(f)|^2}{e_n^2(f)} df\right)^2} = \frac{1}{4\int_0^\infty \frac{|s(f)|^2}{e_n^2(f)} df} = \frac{1}{4\int_0^\infty \frac{|S_V(f)|^2 |p(f)|^2}{e_n^2(f)} df}. \tag{33}$$

Assuming the energy in an event is deposited all at once, $P(t)$ is a delta function at $t = 0$, and $p(f) = 1$. Using the definition of noise equivalent power as $\text{NEP}(f) \equiv e_n(f)/S_V(f)$, the right-most expression in (33) then gives:

$$\Delta E_{\text{rms}} = \left(\int_0^\infty \frac{4df}{\text{NEP}^2}\right)^{-1/2}. \tag{34}$$

The assumption of instantaneous energy input is often not a good one, so the more general form in (33) is also useful.

Figure 9 shows the pulse in Fig. 2c after application of the optimal filter. Note that this output pulse is a factor of $\sqrt{r^2+1}$ faster than the pulse from the detector in Fig. 2, where $r$ is the ratio of TFN to Johnson noise at low frequencies discussed in Sect. 2.1. If negative electrothermal feedback were used, the pulse from the detector would be much faster, and the noise spectrum would look quite different, but the output of the optimal filter would be unchanged.

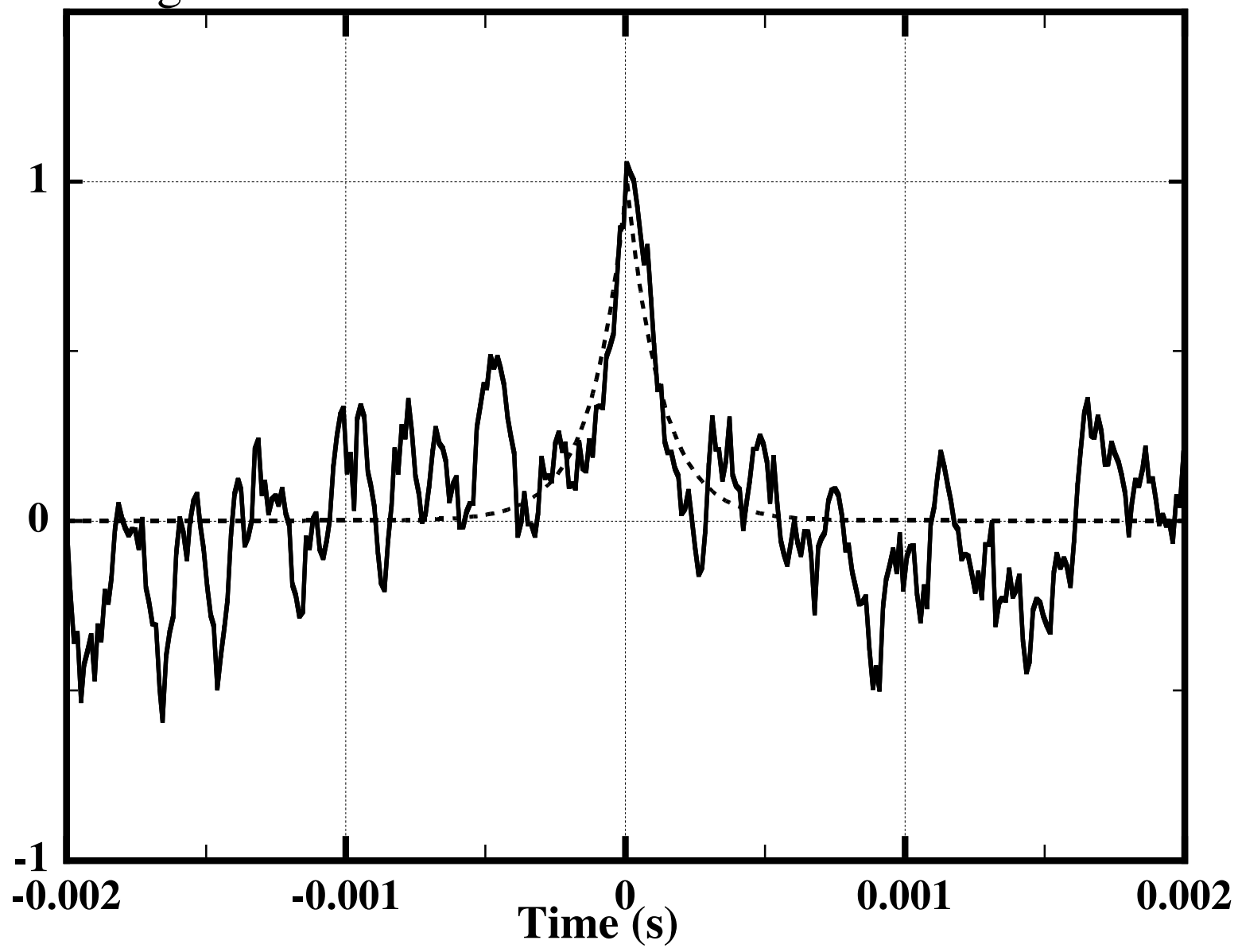

**Fig. 9.** Optimally filtered pulse of Fig. 2c. The r.m.s. noise at the filter output is about 0.22. A noise-free pulse from the same filter is shown by the dashed line

## 2.5 Optimization of Detector and Operating Conditions

Since (34) shows that the energy resolution depends only on *NEP*, we will first calculate the contributions to this by dividing the noise voltage spectral densities derived in Sect. 2.3 by the responsivity $S_V$ from (15). The results are summarized in Table 2.

**Table 2.** *NEP* from major noise sources

Definitions:

$$\alpha \equiv \frac{d\log R}{d\log T} \qquad L_0 \equiv \frac{\alpha P}{GT} \qquad K_L \equiv \frac{R_L}{R_L + R}$$

$$b \equiv \frac{R_L - R}{R_L + R} = 2K_L - 1 \qquad K_F \equiv \frac{1}{1 - bL_0} \cdot \frac{1 + i\omega\tau}{1 + i\omega\tau/(1 - bL_0)}$$

Responsivity (15):

$$S_V(\omega) = \frac{L_0}{I}\frac{1}{1 + i\omega\tau}K_L K_F \quad \text{V/W} \quad \text{(volts/watt)}$$

Thermodynamic Fluctuation Noise (from 21):

$$e_{n\text{-TFN}}(\omega) = \left(4k_B T_0^2 G_0 \cdot \mathrm{F}_{\mathrm{LINK}}(t,\beta)\right)^{1/2} \cdot \frac{L_0}{I}\frac{1}{1 + i\omega\tau}K_L K_F \quad \text{V}/\sqrt{\text{Hz}}$$

$$NEP^2_{\mathrm{TFN}}(\omega) \equiv e^2_{n\text{-TFN}}(\omega)/S_V^2(\omega) = 4k_B T_0^2 G_0 \cdot \mathrm{F}_{\mathrm{LINK}}(t,\beta) \quad \text{W}^2/\text{Hz}$$

Thermistor Johnson Noise (from 22):

$$e_{n\text{J-Therm}}(\omega) = \sqrt{4k_B TR} \cdot K_L K_F$$

$$NEP^2_{\text{J-Therm}}(\omega) = 4k_B TP\left(1 + \omega^2\tau^2\right)/L_0^2$$

Load Resistor Johnson Noise (from 23):

$$e_{n\text{J-Load}}(\omega) = \sqrt{4k_B T_L R_L}\,\frac{R}{R_L}\left(1 + \frac{L_0}{1 + i\omega\tau}\right)K_L K_F$$

$$NEP^2_{\text{J-Load}}(\omega) = 4k_B T_L P\frac{R}{R_L}\frac{\left(1 + L_0\right)^2 + \omega^2\tau^2}{L_0^2}$$

Amplifier Noise (from 25):

$$e_{n\text{-AMP}}(\omega) = \sqrt{e_{nA}^2(\omega) + i_{nA}^2(\omega) \cdot \left(R\left(1 + \frac{L_0}{1 + i\omega\tau}\right)K_L K_F\right)^2}$$

$$NEP^2_{\mathrm{AMP}}(\omega) = e_{nA}^2\, I^2\frac{\left(1 + \omega^2\tau^2\right)}{L_0^2 K_L^2 K_F^2} + i_{nA}^2\, RP\frac{\left(1 + L_0^2\right) + \omega^2\tau^2}{L_0^2}$$

We first notice that the electrothermal feedback and load resistor loading terms $K_F$ and $K_L$ have dropped out of the $NEP$ for both TFN and thermometer Johnson noise. This is as expected, since any noise sources that appear inside a feedback loop or ahead of it will be affected by the feedback in exactly the same way as the signal, leaving the signal to noise ratio and $NEP$ unchanged. The output of the optimal filter is $E_0 |s(f)|^2 / e_n^2(f)$, which is proportional to the square of the signal to noise ratio, so the shape of the pulses is also unchanged by feedback. This means that if $R_L$ is made greater than $R$ to "turn on" negative electrothermal feedback, the pulses at the detector will decay faster because $\tau_e$ is shortened, but the pulses at the output of the optimal filter will be unchanged. On the other hand, if there is a "knob" that allows $\alpha$ to be increased with no electrothermal feedback, the pulses from the detector simply get higher, while at the output of the optimal filter they will get faster.

The amplifier voltage noise however comes in after the feedback loop and so is unaffected by it. The $NEP$ from this then depends inversely on the signal, and so is improved by positive electrothermal feedback ($K_F > 1$) and made worse by negative electrothermal feedback ($K_F < 1$) and load resistor loading ($K_L < 1$). The amplifier current noise goes in ahead of the loop and is modulated by the feedback, so the $K$s again cancel out in the $NEP$. The same is true of load resistor Johnson noise, which when taken as a current noise $i_{nJRL}^2 = 4k_B T_L / R_L$ is entirely equivalent to $i_{nA}^2$. The only dependence on $R_L$ is in $i_{nJRL}^2$ itself.

Optimization depends on its starting assumptions. We will assume that there is some minimum heat sink temperature $T_0$ it is practical to provide, that the detector, due to constraints on desired volume, available materials, and construction methods has some minimum heat capacity $C_0$ at this temperature, and that we have a thermometer technology that offers some maximum sensitivity $\alpha$. In this case, our optimization problem is how to get the best energy resolution for given values of $T_0$, $C_0$, and $\alpha$. The remaining adjustable parameters are $G$, $R$, $R_L$, and $P$, the bias power. The noise sources are independent, making the total $NEP^2$ simply the sum of the individual contributions. $NEP_{\text{TFN}}^2$ and $NEP_{\text{J-Therm}}^2$ do not depend on $R$ and $R_L$, so we are free to choose these to minimize the other contributions. We assume that making $R$ a sufficiently good match to the noise resistance of the amplifier will make this term negligible, and then choose $R_L$ large enough relative to $R$ that we can ignore $NEP_{\text{J-Load}}^2$. This leaves $NEP_{\text{Total}}^2 \approx NEP_{\text{TFN}}^2 + NEP_{\text{J-Therm}}^2$, and $P$ and $G$ as our free parameters.

While we have been assuming a small-signal limit, the temperature increase $t \equiv T/T_0$ produced by the bias power is not necessarily small, and can cause significant changes in the heat capacity of the detector and thermal conductivity of the link. We therefore need a model for these changes. We will assume power-law temperature dependences: $G(tT_0) = G_0 t^\beta$, where $G_0 = G(T_0)$, and $C(tT_0) = C_0 t^\gamma$. Fortunately these are fairly accurate representations for most materials over the required temperature range, with $\beta = 1$ for metals, 3 for insulators, 4 for electron-phonon coupling in metals, and ~5 (at least empirically) for electron-phonon coupling in semiconductors. Heat capacity scaling is $\gamma = 1$ for metals, 3 for insulators, and 0–1 for doped semiconductors. Magnetic materials and superconductors have more complicated behavior, but can usually be approximated by some power law over the range from $T_0$ to $T$. It will be convenient to use $t$ as a parameter rather than $P$, which can now be expressed in terms of $t$ as:

$$G(T)=\frac{dP}{dT} \quad \Rightarrow \quad P=\int_0^P dP=\int_{T_0}^T G(T)\,dT=\int_1^t G_0 t^\beta\, T_0 dt=\frac{G_0 T_0}{\beta+1}\left(t^{\beta+1}-1\right). \qquad (35)$$

We use this expression for $P$ to write $L_0 \equiv \alpha P/(GT)=\alpha\left(1-1/t^{\beta+1}\right)/(\beta+1)$. Substituting the above and $\tau \equiv C/G=C_0 t^{\gamma-\beta}/G_0$ into the expressions for $NEP_{\text{TFN}}^2$ and $NEP_{\text{J-Therm}}^2$ from Table 2, we get

$$NEP_{\text{Total}}^2(\omega)=\left[4k_B T_0^2 G_0\right]\cdot\left(\mathrm{F}_{\text{LINK}}(t,\beta)+\frac{(\beta+1)t^{2\beta+3}}{\alpha^2\left(t^{\beta+1}-1\right)}\left(1+\omega^2\tau^2\right)\right). \qquad (36)$$

This is readily integrated in (34) to give the expected energy resolution:

$$\Delta E_{\text{RMS}}=\left(\left[k_B T_0^2 C_0\right]\frac{4(\beta+1)t^{\gamma+2}}{\alpha^2\left(1-t^{-(\beta+1)}\right)}\sqrt{1+\frac{\alpha^2\left(t^{\beta+1}-1\right)\mathrm{F}_{\text{LINK}}(t,\beta)}{(\beta+1)t^{2\beta+3}}}\right)^{1/2} \equiv \xi(t,\alpha,\beta,\gamma)\sqrt{k_B T_0^2 C_0}. \quad (37)$$

Note that $G$, the thermal conductivity of the link, does not appear in this expression, although its temperature dependence does. This means that we are free to choose $G$ to give any desired time constant (but see Sects. 2.8 and 2.10 for limits to the validity of this statement for real detectors and thermometers). We can see that for $\alpha$ sufficiently large to neglect the 1 under the radical, $\Delta E \propto \sqrt{k_B T_0^2 C_0/\alpha}$.

We have only the bias power available for optimization, so we insert the appropriate function for $\mathrm{F}_{\text{LINK}}$, which would be from (18) for a hot-electron device or one with a very specular conducting link and from (19) for a perfectly diffusive link, and adjust $t$ to minimize $\xi$. Figure 10 shows the variation of $\xi$ with $t$ for various values of $\alpha$, $\beta$, and $\gamma$. The dependence of $\xi_{\min}$ on $\beta$ and $\gamma$ is weak, although $\gamma$ significantly affects the optimum value of $t$. Figure 2 of [5] gives $t_{\text{opt}}$ and $\xi_{\min}$ as functions of $\alpha$ for some values of $\beta$ and $\gamma$.

For power detectors, the figure of merit is *NEP* rather than $\Delta E$, and the form of (36) makes it appear that this depends on $G$ rather than $C$. Mather, however, argues in [11] that the proper figure of merit is always the *NEP* at some chosen frequency $\omega$, and writes $NEP^2(\omega)=\left[4k_B T_0^2 C_0\omega\right]\cdot \mathrm{f}(t,g,q,\alpha,\beta,\gamma)$, where $g$ and $q$ are respectively the normalized link conductivity and background power. Conductivity is assumed to be a free parameter, so f can be minimized with respect to both $t$ and $g$. The *NEP* optimized for any given frequency then has qualitatively the same dependences as $\Delta E$, with $\mathrm{f} \propto \alpha^{-1/2}$ for sufficiently large $\alpha$.

## 2.6 Complex Load Impedance

The derivations above do not require that the load resistance, $R_L$, be real and we are free to include intentional or stray capacitive and inductive effects in its value. With the bias circuit of Fig. 4a, the high end of the load resistor is fixed, and capacitance in parallel with the detector is equivalent to a parallel capacitance on the load resistor and can be accounted for in this fashion. This is not true when a time varying signal is applied to the top of the load resistor (as in Sect. 2.7 below). In this case, and to handle an inductor in series with the detector, it would be necessary to redo the analysis to explicitly include these components (but see Sect. 2.9). For high impedance thermometers, such as doped semiconductors, stray capacitance effects usually dominate, while for very low impedance thermometers such as superconducting TES, only stray inductance is normally important.

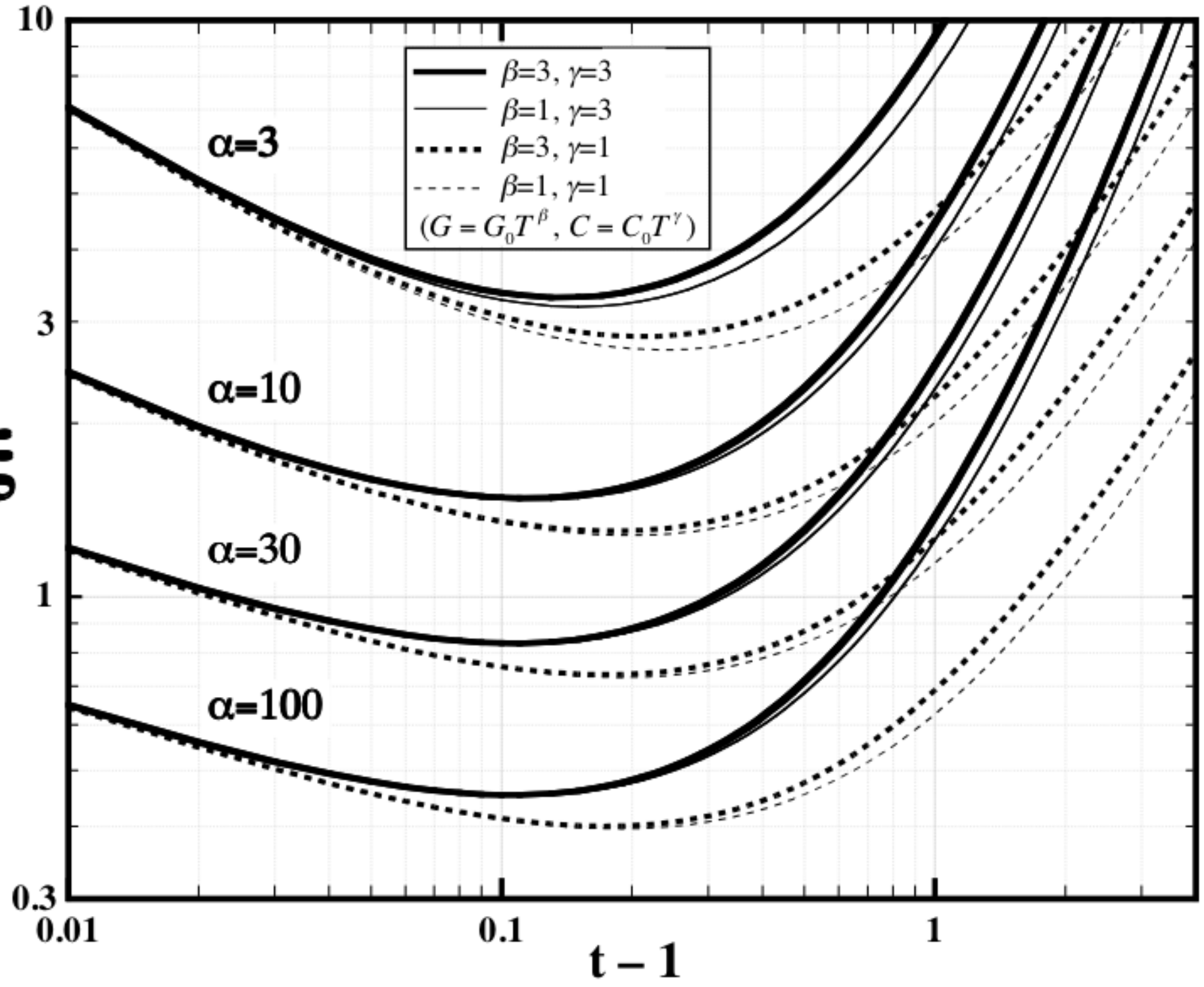

**Fig. 10.** Normalized energy resolution as a function of the temperature increase $t-1=\Delta T_{\mathrm{Bias}}/T_0$ produced by bias power. The effective link temperature function $F_{\mathrm{LINK}}$ appropriate for radiative transfer was used, but the result is not appreciably different for the diffuse scattering case.

### 2.7 External electrothermal feedback

The practical advantages of electrothermal feedback, and particularly of large amounts of negative ETF, were discussed above and will come up again in Sect. 3. Large feedback effects require the absolute value of the loop gain, $\beta A = bL_0/(1+i\omega t)$, to be $>>1$ at low frequencies. For $R_L>0$, $b\equiv(R_L-R)/(R_L+R)$ will always lie between –1 and +1, so this would normally require $|L_0|>>1$, which can only be obtained with extremely sensitive thermometers.

It is possible however to get arbitrarily large feedback effects with low-$\alpha$ thermometers by using the gain of the electronic amplifier. Suppose the thermometer voltage V is amplified by a factor $\lambda$ and then added to the bias supply voltage $V_B$ at the top of $R_L$. We can then write the current as $I = (V_B + \lambda V - V)/R_L$, and $dI/dV = (\lambda - 1)/R_L$. All of the equations derived in Sect. 2 depend on $R_L$ only through $dI/dV$, so we take this external feedback into account simply by substituting an effective value of $R_L$ equal to $R_L/(1-\lambda)$ for all occurrences of $R_L$ in $K_F$ and $K_L$ (the explicit appearance of $R_L$ in the expression for load resistor Johnson noise is only to give the magnitude of the Johnson noise current source $i_{nJRL} = \sqrt{4k_B T_L/R_L}$, which is not affected by feedback). In particular the expression for *b* will have a zero in the denominator for $\lambda = (R_L + R)/R$, so *b* and $\beta A$ can become arbitrarily large positive or negative with $\lambda$ sufficiently close to this critical value.

All of the equations derived above are then valid for this situation with the exception of the one for amplifier voltage noise. Since the amplifier noise is now inside the feedback loop, it is also modified by the feedback, and in principle one can get the benefits of negative ETF without the potentially adverse effect on $NEP_{\mathrm{AMP}}$ produced by intrinsic negative ETF. Of course, improvements in $NEP_{J-Therm}$ and energy resolution depend on $L_0$ and not on the feedback, and come only from increasing the thermometer sensitivity. Also, when *b* is large it becomes very sensitive to changes in *R*, so problems with stability in the face of large signals or changing background power will limit the practical amount of external feedback.

## 2.8 Thermometer nonlinearities

Up to this point we have assumed that the thermometer resistance depends only on *T*, *i.e.* $R = R(T)$. More generally, $R = R(T,V)$. This behavior is normally referred to as "nonlinearity", since it makes ohm's law a nonlinear equation. For small signals, however, it can be linearized around the operating point and characterized by the two partial derivatives

$$\alpha_V \equiv \left.\frac{\partial \log R}{\partial \log T}\right)_V \quad \text{and} \quad \beta_V \equiv \left.\frac{\partial \log R}{\partial \log V}\right)_T . \tag{38}$$

The following chapters discuss specifics of the nonlinear behavior for the most common types of resistive thermometers. It is straightforward however to incorporate the linearized form into the relations derived so far. This was first done for $S_V(\omega)$ and $Z(\omega)$ by Mather [13].

For example, to re-derive the D.C. response in (8) from (2), we must find $dR/dT$ when $R = R(T,V)$. We can write

$$dR = \left.\frac{\partial R}{\partial T}\right)_V dT + \left.\frac{\partial R}{\partial V}\right)_T dV \quad \Rightarrow \quad \frac{dR}{dT} = \frac{R}{T}\alpha_V + \frac{R}{V}\beta_V \frac{dV}{dR}\frac{dR}{dT} . \tag{39}$$

Using $dV/dR$ from (3), this can be solved for $dR/dT$:

$$\frac{dR}{dT} = \frac{R}{T}\alpha_V \cdot \left(1 - \beta_V \frac{R_L}{R_L + R}\right)^{-1}. \tag{40}$$

Substituting this into (2) with the other terms the same as for (8) gives

$$\frac{dV}{dP_{in}} = \frac{L_V}{I(1 + i\omega\tau)} \cdot \frac{1}{(1 - K_L\beta_V)} K_L, \tag{41}$$

where we have changed the subscript on $L_0$ as a reminder that it is now $L_V \equiv \alpha_V P/GT$. Next, we can easily compute the changes to the loop gain $\beta A$ that controls feedback effects, since $\beta$ from (10) remains the same, and the gain $A$ in (41) differs from (8) only by the additional term $1/(1 - K_L\beta_V)$. (Note that the feedback factor $\beta$ is unrelated to $\beta_V$.) If we still want to write $\beta A$ as $b_V L_V$, then $b$ must be modified to $b_V = b/(1 - K_L\beta_V)$. As $\beta_V$ is usually negative, the new term is degenerative and can significantly reduce both gain and the effects of feedback.

This process can be continued for the response to major noise sources with the results shown in Table 3. Note that the responsivity and major noise terms all include the added factor of $1/(1 - K_{L\text{-}V}\beta_V)$ and the modified $K_{F\text{-}V}$. These will therefore cancel out in the *NEP* calculation, leaving this and the energy resolution unchanged. A reservation must be made about this regarding thermistor Johnson noise. Johnson noise and the nonlinearity are both internal to the thermistor. The result in (45) was derived assuming that the Johnson noise voltage modulated its own resistance, but the physics behind the nonlinearity is unspecified and it is not obvious that this should always be the case.

## 2.9 Voltage Output vs Current Output

We have pointed several advantages of negative electrothermal feedback, and one often wishes to maximize it. Since the feedback factor is $1/(1 - bL_0)$, we then want $bL_0$ negative and as large as possible. For $\alpha < 0$, $L_0$ is negative, so we need $b = (R_L - R)/(R_L + R)$ close to +1. This is readily done by making $R_L >> R$, since this also minimizes its loading effects ($K_L \approx 1$) and its Johnson noise contribution to the *NEP*. However, for positive temperature coefficient thermometers, we want $b$ close to –1, and this requires $R_L << R$. We would then have significant load resistor Johnson noise, and the signal level would be reduced to a point where amplifier noise is important.

The obvious solution in the latter situation is to switch to using the current through the thermistor rather than the voltage across it as the output signal. Signal loading effects are then minimized for $R_L << R$, and it is easily shown that this condition also makes load resistor Johnson noise insignificant.

Rather than re-derive all of the relations in Sect. 2 for current output, we can make use of the simple dual circuit transformation to change them to this form. This allows us to switch all of the *V*'s for *I*'s (and *I*'s for *V*'s) if accompanied by the other substitutions

**Table 3.** Response with thermometer nonlinearity: $R = R(T, V)$

Definitions:

$$\alpha_V \equiv \left.\frac{\partial \log R}{\partial \log T}\right)_V \qquad \beta_V \equiv \left.\frac{\partial \log R}{\partial \log V}\right)_T \qquad L_V \equiv \frac{\alpha_V P}{GT} \qquad K_{L\text{-}V} \equiv \frac{R_L}{R_L + R}$$

$$b_V \equiv \frac{R_L - R}{R_L(1-\beta_V) + R} \qquad K_{F\text{-}V} \equiv \frac{1}{1 - b_V L_V} \cdot \frac{1 + i\omega\tau}{1 + i\omega\tau/(1 - b_V L_V)}$$

Responsivity:

$$S_V(\omega) = \frac{L_V}{I(1 - K_{L\text{-}V}\beta_V)} \cdot \frac{1}{1 + i\omega\tau} K_{L\text{-}V} K_{F\text{-}V} \qquad (42)$$

Impedance:

$$Z(\omega) \equiv \frac{dV}{dI} = \frac{(1 + i\omega\tau) + L_V}{(1 + i\omega\tau)(1 - \beta_V) - L_V} \cdot R \qquad (43)$$

Thermodynamic Fluctuation Noise:

$$e_{n\text{-TFN}}(\omega) = \left(4 k_B T_0^2 G_0 \cdot \mathrm{F}_{\mathrm{LINK}}(t, \beta)\right)^{1/2} \frac{L_V}{I(1 - K_{L\text{-}V}\beta_V)} \cdot \frac{1}{1 + i\omega\tau} K_{L\text{-}V} K_{F\text{-}V} \qquad (44)$$

Thermistor Johnson Noise:

$$e_{n\mathrm{J\text{-}Therm}}(\omega) = \sqrt{4 k_B T R}\, \frac{1}{(1 - K_{L\text{-}V}\beta_V)} K_{L\text{-}V} K_{F\text{-}V} \qquad (45)$$

Load Resistor Johnson Noise:

$$e_{n\mathrm{J\text{-}Load}}(\omega) = \sqrt{4 k_B T_L R_L}\, \frac{R}{R_L}\left(1 + \frac{L_V}{1 + i\omega\tau}\right) \frac{1}{(1 - K_{L\text{-}V}\beta_V)} K_{L\text{-}V} K_{F\text{-}V} \qquad (46)$$

Amplifier Noise:

$$e_{n\text{-AMP}}(\omega) = \sqrt{e_{nA}^2(\omega) + i_{nA}^2(\omega) \cdot \left(\frac{R}{(1 - K_{L\text{-}V}\beta_V)}\left(1 + \frac{L_V}{1 + i\omega\tau}\right) K_{L\text{-}V} K_{F\text{-}V}\right)^2} \qquad (47)$$

given in Table 4. It will also work for more complicated circuits that incorporate stray capacitance and inductance. The circuit transformation is usually obvious in our simple cases. If the effects of voltage amplifier parallel input capacitance have been included, for instance, this would just become the series inductance of the current amplifier input in the transformed equation. For more complex circuits, one should consult the full algorithm for the transformation under "Duality" in a suitable textbook [14].

**Table 4.** Dual circuit transformations

| | | |
|---|---|---|
| $V$ (across a component) | → | $I$ (through transformed component) |
| $I$ (through a component) | → | $V$ (across transformed component) |
| $R$ (resistance) | → | $1/R$ (conductance) |
| $C$ (parallel) | → | $L$ (series) |
| $L$ (series) | → | $C$ (parallel) |

The results for all the formulae of this section are given in Table 5. Only the versions including thermistor nonlinearity are given; to recover the linear form, set $\beta_I = 0$. Nonlinear thermistors are characterized by partial derivatives of $R(T,I)$ rather than $R(T,V)$. The *NEP* contributions from TFN and thermistor Johnson noise are unchanged. Those from the load resistor and amplifier are given in the table. These and other useful current-output formulae are derived directly in [15].

## 2.10 Complications

The detector of Fig. 1, where the absorber, thermometer, and structure are regarded as a single isothermal entity, is clearly an idealization, as is the supposition that deposited energy in an event is instantaneously in equilibrium in all available channels. In this section we discuss the effects of some of the more important complications.

***Internal time constants and internal fluctuation noise.*** A detector can more reasonably be modeled as separate components: a thermometer, an absorber, and perhaps some structure, connected by thermal resistances. In addition, any component with more than one channel for energy content, such as a metal with both lattice phonons and conduction electrons, should be modeled as two or more parts with their own heat capacities connected by appropriate thermal conductances. This has two potentially major effects, which we illustrate by a simple (but realistic) example where the energy-absorbing part of the detector has a finite thermal conductivity, $G_A$, to the rest of the detector, including the thermometer. When an event deposits energy in the absorber, the thermometer temperature will have a finite risetime, limited by $G_A$ and the heat capacities of the absorber and thermometer. The qualitative effect of this is apparent by looking at Fig. 2b. A better quantitative impression can be obtained from Fig. 3, where $G_A$ introduces another pole in the signal response. Above the frequency of this second pole, the signal to TFN ratio, which is otherwise completely independent of frequency, will start to drop as $1/f$. This is the same effect that otherwise occurs above the frequency where the TFN drops below the level of the Johnson noise, so it will not seriously impact the energy resolution if the new pole is at a considerably higher frequency than this crossing point. The crossing, where (21) and (22) are equal, is at $\sim \alpha/4$ times the frequency of the main pole of the signal, so this is an increasingly stringent requirement

**Table 5.** Equations for current output with $R = R(T, I)$

Dual Transforms and Definitions:

$$\alpha_V \rightarrow -\alpha_I, \qquad \alpha_I \equiv \left.\frac{\partial \log R}{\partial \log T}\right)_I; \qquad \beta_V \rightarrow -\beta_I, \qquad \beta_I \equiv \left.\frac{\partial \log R}{\partial \log I}\right)_T$$

$$K_{L\text{-}V} \rightarrow 1 - K_{L\text{-}V} = K_{L\text{-}I} \equiv \frac{R}{R_L + R}; \qquad L_V \rightarrow -L_I, \qquad L_I \equiv \frac{\alpha_I P}{GT}$$

$$b_V \equiv \frac{b}{\left(1 - K_{L\text{-}V}\beta_V\right)} \rightarrow -b_I, \qquad b_I \equiv \frac{b}{\left(1 + K_{L\text{-}I}\beta_I\right)}; \qquad b \equiv \frac{R_L - R}{R_L + R}.$$

$$K_{F\text{-}V} \equiv \frac{1}{1 - b_V L_V} \cdot \frac{1 + i\omega\tau}{1 + i\omega\tau / \left(1 - b_V L_V\right)} \rightarrow \frac{1}{1 - b_I L_I} \cdot \frac{1 + i\omega\tau}{1 + i\omega\tau / \left(1 - b_I L_I\right)} \equiv K_{F\text{-}I}$$

Responsivity – from (42):

$$S_I(\omega) = \frac{-L_I}{V\left(1 + K_{L\text{-}I}\beta_I\right)} \cdot \frac{1}{1 + i\omega\tau} K_{L\text{-}I} K_{F\text{-}I} \tag{48}$$

Admittance – from (43):

$$A(\omega) \equiv \frac{dI}{dV} = \frac{\left(1 + i\omega\tau\right) - L_I}{\left(1 + i\omega\tau\right)\left(1 + \beta_I\right) + L_I} \cdot \frac{1}{R} \tag{49}$$

Thermodynamic Fluctuation Noise – from (44):

$$i_{n\text{-TFN}}(\omega) = \left(4k_B T_0^2 G_0 \cdot \mathrm{F}_{\mathrm{LINK}}(t, \beta)\right)^{1/2} \cdot \frac{L_I}{V\left(1 + K_{L\text{-}I}\beta_I\right)} \frac{1}{1 + i\omega\tau} K_{L\text{-}I} K_{F\text{-}I} \tag{50}$$

Thermistor Johnson Noise – from (45):

$$i_{n\text{J-Therm}}(\omega) = \sqrt{4k_B T / R} \cdot \frac{1}{\left(1 + K_{L\text{-}I}\beta_I\right)} K_{L\text{-}I} K_{F\text{-}I} \tag{51}$$

Load Resistor Johnson Noise – from (46):

$$i_{n\text{J-Load}}(\omega) = \sqrt{4k_B T_L / R_L}\, \frac{R_L}{R}\left(1 - \frac{L_I}{1 + i\omega\tau}\right) \frac{1}{\left(1 - K_{L\text{-}I}\beta_I\right)} K_{L\text{-}I} K_{F\text{-}I} \tag{52}$$

$$NEP^2_{\text{J-Load}}(\omega) = 4k_B T_L P \frac{R_L}{R} \frac{\left(1 - L_I\right)^2 + \omega^2\tau^2}{L_I^2} \tag{53}$$

Amplifier Noise – from (47):

$$i_{n\text{-AMP}}(\omega) = \sqrt{i_{nA}^2(\omega) + e_{nA}^2(\omega) \cdot \left(\frac{1}{R\left(1 + K_{L\text{-}I}\beta_I\right)}\left(1 - \frac{L_I}{1 + i\omega\tau}\right) K_{L\text{-}I} K_{F\text{-}I}\right)^2} \tag{54}$$

$$NEP^2_{\text{AMP}}(\omega) = i_{nA}^2\, V^2 \frac{\left(1 + K_{L\text{-}I}\beta_I\right)^2\left(1 + \omega^2\tau^2\right)}{L_I^2 K_{L\text{-}I}^2 K_{F\text{-}I}^2} + e_{nA}^2 \frac{P}{R} \frac{\left(1 + L_I^2\right) + \omega^2\tau^2}{L_I^2} \tag{55}$$

as the thermometer sensitivity becomes very high. Put another way, to get the improvement in energy resolution expected from an increase in thermometer sensitivity, you must ensure that internal time constants in the detector are of order $\alpha$ times shorter than the main detector time constant.

The second detrimental effect is that random exchange of energy between parts of the detector can produce additional thermodynamic fluctuation noise. This "internal TFN" is another term that must be added to the total *NEP*. Reference [9] derives the proper equations for a few models of detector internal structure that are reasonable representations of some real detectors. These include both the altered frequency response to signals and the internal TFN. A general approach is outlined that allows algebraic equations to be derived for other internal structures, but these quickly become unwieldy. A general matrix method is demonstrated in [15] that allows numerical solutions to be obtained for very complex structures.

There is one additional effect of internal thermal resistances. If the thermometer itself is isolated from the rest of the detector, as for instance by the electron-phonon coupling in a thermistor where the resistance depends on the electron temperature, then the bias power dissipated in the thermistor will create a temperature drop across the coupling resistance. If the temperature coefficient $\beta$ is large, and it is 4 – 5 for electron-lattice coupling, then as this temperature drop increases the thermometer temperature quickly becomes insensitive to the external temperature. This loss of effective thermometer sensitivity is a particularly severe problem for doped semiconductor thermistors, as discussed in the next chapter. Models worked out in [10] include this effect, as well as the increased Johnson noise, additional internal TFN, and loss of high frequency signal response that go with it.

***Thermalization noise.*** We have been assuming that all of the event energy comes instantaneously into equilibrium with at least the absorber portion of the detector. It is possible however for part of the energy to go into a channel that is so weakly coupled with the others that it takes a significant length of time to come into equilibrium with them. For example, when an X-ray photon is absorbed in a semiconductor about one-third of the energy initially goes into producing electron-hole pairs. At low temperatures, these are mostly trapped on impurity sites, and recombinations that return this energy to the phonon system can be very long. If the equilibration time is not short compared with $\tau/\alpha$, then it will have an effect on detector performance.

There are three forms of such effects. The first is simply due to part of the energy being completely lost, or slowed in delivery relative to the main thermal time constant so that it is measured with poorer signal to noise ratio. These change the optimal filter and the energy scale to make the resolution worse, but are still independent of event energy. This would not usually be regarded as thermalization "noise", although it involves a loss of resolution related to thermalization processes. To some extent it is completely unavoidable: thermalization in a phonon system cannot be faster than the sound crossing time of the detector, for instance. For effects like this where the time is not too long, the solution is to slow the detector by making $G$ smaller, increasing $\tau$ until $\tau/\alpha$ is at least comparable to the thermalization time. Twerenbold [16] includes a basic discussion of the physics involved in this process.

The other forms of thermalization effects are ones that vary from one event to the next. The variations can be either statistical, such as the Fano fluctuations in the fraction of energy going into ionization, or position-dependent. Position dependence can come from detector geometry or from local variations in a parameter like defect density. This kind of "noise" generally causes a loss of energy resolution that is proportional to the event energy. Position dependence produced by variations in the energy collection time can be reduced by making the detector slower if the application allows it, or by altering the filtering to compromise between signal-to-noise ratio and insensitivity to risetime.

## 3 Limitations of Linear Theory

The linear theory of Sect. 2 is simple and allows many useful generalizations, but the unfortunate truth is that many thermal detectors, particularly the small ones applied to high-resolution X-ray measurements, are often run well into the nonlinear regime. As shown in (37), the fundamental parameters determining energy resolution are the bath temperature and detector heat capacity, and it is technologically practical to make detectors for soft X rays with such small heat capacities that $\Delta T$ is a substantial fraction of $T_0$ when a photon is absorbed. (Detectors for higher energy photons and other things with small cross sections usually require such a massive absorber than nonlinearity is not a problem). All the thermodynamic parameters tend to change rapidly with temperature, and thermometer characteristics can be the worst offender. The very sensitive and promising superconducting transition edge sensor (TES) represents the extreme case. Its resistance can go from zero to the full normal value with less than 1% temperature change. Within this range, $\alpha$ is changing rapidly, and outside of it the thermometer is completely saturated. Modest amounts of nonlinearity can be handled simply by careful calibration of the pulse height vs. energy relation. However, significant increases in temperature will change the noise characteristics during the pulse, leading to correlations between the noise at different frequencies and invalidating the optimal filter as calculated in Sect. 2.4.

One standard way of handling this is to increase the heat capacity of the detector until the nonlinearity is acceptable for the highest event energy of interest, $E_{\text{MAX}}$. This has the interesting consequence that the energy resolution becomes independent of thermometer sensitivity. If the width of the superconducting-to-normal transition is narrowed, $\alpha$ is approximately proportional to the inverse of the width, and the heat capacity must be similarly increased to keep the event within the same part of the transition. The energy resolution, which scales approximately as $(C/\alpha)^{1/2}$, is unchanged by this, but is now proportional to $E_{\text{MAX}}^{1/2}$. There is still a major benefit to increasing $\alpha$, however. Since $C$ is to be increased above the minimum technologically feasible value, this extra "budget" allows a wider choice of materials to be used for constructing the detector that can provide faster and more complete thermalization or better thermal contact. This makes the detector a better approximation of an ideal device, particularly if short time constants are desired.

Fixsen et al. [17,18] have taken another approach and developed an optimal filtering algorithm that properly accounts for the noise correlations between frequencies. They find that for any amount of nonlinearity, including complete saturation of the TES, the resulting resolution is no worse than it would have been had $C$ been increased to lessen the nonlinear effects. This is a very important result, since it means that much better resolution can be obtained at all lower energies with the same detector. The signal processing is more complex, but could be worthwhile for an application that covers a wide range of event energies. Some of these authors have proposed that the best design would be to choose a $C$ that just starts to saturate completely at $E_{\mathrm{MAX}}$. This is at least a factor of three lower heat capacity than would normally be required to keep nonlinearity to a moderate level, and results in a significant resolution improvement at low energies. Heat capacity could be reduced still further without hurting the resolution at $E_{\mathrm{MAX}}$, but with the detector fully saturated there is no way to detect an accidental additional event that adds to the total energy. Somewhat below saturation, the pulse shape distortion produced by the pileup can still be detected and used to reject the event. For applications where the best energy resolution over a very wide dynamic range is important, it might also be practical to include a second thermometer with lower sensitivity for the sole purpose of detecting pileup while the main thermometer is saturated.

Up to this point, we have been considering isolated events, but the large nonlinearities of thermal detectors can make pileup a much more severe problem than it is in the usually rather linear ionization detectors. For the small-signal linear case, pileup considerations are the same as they are for charge detectors. The only thing that matters is the output of the filter, and with a sensitive thermometer the signal-to-noise ratio is good at frequencies far above the thermal corner at $f_{\mathrm{c}} = G/(2\pi C)$. Therefore even the optimal filter can show little pileup at event separations much smaller than the thermal time constant, and other filters can be used that allow even higher rates with some compromise in energy resolution.

For pulses big enough to show nonlinear effects, however, a second event that arrives before the detector has actually cooled sufficiently will show a different response than the first event. For large events, this could require several thermal time constants. Here negative electrothermal feedback, either intrinsic or external, can make a big difference. As can be seen in (15), feedback will speed the actual cooling rate by a factor of $(1 - bL_0)$, allowing counting rates to increase by a similar amount before pileup effects become important. This points out a drawback of the scheme proposed above where large events are allowed to completely saturate the thermometer, since feedback then quits operating and thermal recovery times become much longer. How serious this is depends on the application. For many astronomical observations, for instance, the rate of high energy events is very small compared to that at lower energies, so the increased deadtime from these might be a small price to pay for significantly improved resolution at low energies.

# 4 Other Thermometer Types

There are many other potential thermometer systems. Besides the reactive analogs of thermistors ($L(T)$ and $C(T)$), we have pyro-electric ($V(T)$) and paramagnetic ($M(T)$) devices. Any temperature-dependent physical parameter that can be measured is a potential thermometer, but the only one I know of in addition to the above list that has been used for low temperature detectors is the tunneling current of an NIS junction. It would be handy to make some grand synthesis of all thermometer types so they could be compared directly, but in practice qualitative and quantitative differences in the operation change the approximations that can be used and often require entirely different figures of merit. The only universal quantity is the thermodynamic fluctuation noise. As we have seen, this does not itself limit the energy resolution, but it does set a scale where the resolution is proportional to the square root of the bandwidth over which the signal to noise ratio is approximately constant. This frequency range is where TFN is the dominant noise source and the signal frequency response is dominated by the same thermal time constant that determines the TFN spectrum. For resistive thermometers, the Johnson noise provided a fundamental and in most cases attainable limit to how large this bandwidth could be. For other thermometer types, the Johnson noise may be negligible, and some other noise source or attenuation of the signal frequency response will become the limiting factor.

The reactive thermometers appear to be much like thermistors. They have an output that depends on both their sensitivity $\alpha \equiv d\log X / d\log T$ and the bias current, but the dissipation is only $1/Q$ of an equivalent resistive element. Here $Q$ is the standard "Quality Factor" for a reactive element, among whose equivalent definitions is $Q \equiv$ (energy stored energy per cycle)/(energy dissipated per cycle). The optimum bias current is therefore increased by $Q^{1/2}$ and the Johnson noise is reduced by $Q^{1/2}$. This improves the TFN to Johnson noise ratio by a factor of $Q$, so we should be able to define an effective sensitivity "$\alpha_{\mathrm{Eff}}$" $\equiv Q \cdot \alpha$ and take over most of the thermistor results. Although most of these devices have relatively low $\alpha$'s, available $Q$'s can be very high (often $10^5$–$10^6$). Unfortunately, something else usually dominates the noise or the signal has another pole before the TFN drops below the level of the Johnson noise. In these cases $\alpha_{\mathrm{Eff}}$ is not a good figure of merit, and the optimization will probably look quite different than it does for a resistive thermometer. So each of these new thermometer types must be analyzed individually, at least until it is determined whether it is amplifier noise, internal time constants, or both, or something entirely different, that dominates the behavior. The result for the optimal filter, at least in the form of (33), should be generally applicable for any detector in the small-signal limit.

### 4.1 Kinetic Inductance

The kinetic inductance thermometer uses the steep temperature variation in magnetic penetration depth in a superconductor just below its transition temperature to modulate the inductance of a nearby coil [19,20]. Usually an insulating film is deposited on top of a thin superconducting plane, and the coil is deposited on top of that. The detector is well matched to a SQUID amplifier, and in principle could be very sensitive, with Q values close to $10^6$. Versions have recently been proposed that construct an array of pixels with resonators of different frequencies, allowing the entire array to be read out in parallel with multi-frequency microwave excitation. This requires no physical connections to the detector array, and puts almost all of the readout electronics at room temperature [21]. Detectors of this type developed thus far operate far from equilibrium.

## 4.2 Magnetic

The net magnetization of a system of spins of moment $\mu$ in a magnetic field depends on temperature as $\tanh(\mu B/kT)$. The spin system must be made dilute enough to avoid strong spin-spin interactions, which severely limits the sensitivity per unit volume. However, tiny changes in magnetization can sensed with a pickup coil and SQUID amplifier, and no excitation is required. This means that only third-order effects produce dissipation from the readout system, and the power input can be tiny [22]. The system has almost no intrinsic Johnson noise, so the useful bandwidth is limited instead by some combination of the amplifier noise, internal thermodynamic fluctuation noise, and internal coupling times that drop off the signal relative to the TFN at high frequencies.

With a magnetic thermometer, the signal amplitude is approximately proportional to the thermometer volume. When the amplifier noise dominates, an optimum design will then enlarge the thermometer until its heat capacity is equal to that of the rest of the detector. This changes the scaling of energy resolution vs. absorber $C$, and makes signal level per unit heat capacity an important figure of merit for the thermometer. If the amplifier and its coupling to the field are good enough, the limiting factor will be the internal TFN and signal roll-off due to internal time constants. The most fundamental of these is the coupling time between the spin system and the lattice, which for high resolution should be much smaller than the detector time constant. The very small power dissipation would allow small values for $G$ and large thermal time constants, but application requirements and technical feasibility put a limit on this, so another fundamental figure of merit for magnetic thermometers is the spin-lattice relaxation time. This is normally quite long in dielectric materials, so metallic systems are favored. As different aspects of these detectors are improved, the optimization and figures of merit will change.

Magnetic thermometers have received much less attention than thermistors up to this point, but they have recently attained resolution levels comparable to the best thermistor results [23]. Since their different optimization gives them additional advantages in certain applications, they now appear very promising and are discussed in much more detail in a subsequent chapter in this volume.

## 4.3 NIS junctions

These are tunnel junctions with a superconductor on one side and a normal metal on the other. They are biased so that only electrons in the high-energy tail of the thermal distribution in the normal metal can tunnel, making the tunneling current extremely sensitive to temperature [24]. These devices can also be designed to refrigerate themselves, and there is some interest in using them as “microcoolers” for other detectors [25].

## 4.4 Pyro-electric

Like the magnetic thermometer, this sensor requires no excitation. For an all-metallic system, it should be possible to make an extremely fast detector. The figures of merit and optimization are quite different from most other sensors, which leads to the possibility that there are applications for which it might have an advantage. However its basic properties do not seem to favor it for very high energy resolution [26].

A number of people read early drafts of this manuscript and made helpful comments. I thank in particular Kent Irwin, Lindsay Rocks, Yoh Takei, and John Vaillancourt for their thorough reviews and thoughtful suggestions that have greatly improved the final version. This work was supported in part by NASA grant NAG5-5404.